\documentclass[review]{elsarticle}
\usepackage{hyperref}
\usepackage{amsmath,amssymb,amsfonts}
\usepackage{float}
\usepackage{verbatim} 
\usepackage{apalike}
\usepackage{graphicx}
\usepackage{textcomp}
\usepackage[table,xcdraw]{xcolor}
\usepackage{todonotes}
\usepackage{booktabs}
\usepackage{multirow}
\usepackage{xcolor}
\usepackage{hhline}
\usepackage{subcaption}
\usepackage{hyperref}
\usepackage{titlesec}
\usepackage{diagbox}
\usepackage{enumitem}
\restylefloat{figure}
\restylefloat{table}

\newcommand{\review}[1]{\textcolor{black}{#1}}
\newcommand{\reviewtwo}[1]{\textcolor{black}{#1}}
\newcommand{\reviewthree}[1]{\textcolor{black}{#1}}
\let\cite\citep

\journal{Expert Systems with Applications}

\biboptions{authoryear}

\begin{document}
\begin{frontmatter}

\begin{titlepage}
\begin{center}
\vspace*{1cm}

\textbf{ \large Fast \& Furious: Modelling Malware\\Detection as Evolving Data Streams}

\vspace{1.5cm}

Fabrício Ceschin$^{a}$ (fjoceschin@inf.ufpr.br), Marcus Botacin$^a$ (mfbotacin@inf.ufpr.br), Heitor Murilo	Gomes$^b$ (heitor.gomes@vuw.ac.nz), Felipe Pinagé$^a$ (fapinage@inf.ufpr.br), Luiz S. Oliveira$^a$ (lesoliveira@inf.ufpr.br), André Grégio$^a$ (gregio@inf.ufpr.br) \\

\hspace{10pt}

\begin{flushleft}
\small  
$^a$ Department of Informatics, Federal University of Paraná -- Curitiba, Brazil \\
$^b$ School of Engineering and Computer Science, Victoria University of Wellington -- Wellington, New Zealand \\

\vspace{1cm}
\textbf{Corresponding Author:} \\
Fabrício Ceschin \\
Rua Cel. Francisco Heráclito dos Santos, 100 -- Curitiba, Paraná, Brazil \\
Tel: +55 (41) 99658-8299 \\
Email: fjoceschin@inf.ufpr.br

\end{flushleft}        
\end{center}
\end{titlepage}

\title{Fast \& Furious: On the Modelling of Malware Detection as an Evolving Data Stream}

\author[label1]{Fabrício Ceschin}
\ead{fjoceschin@inf.ufpr.br}

\author[label1]{Marcus Botacin}
\ead{mfbotacin@inf.ufpr.br}

\author[label2]{Heitor Murilo Gomes}
\ead{heitor.gomes@vuw.ac.nz}

\author[label1]{Felipe Pinagé}
\ead{fapinage@inf.ufpr.br}

\author[label1]{Luiz S. Oliveira}
\ead{lesoliveira@inf.ufpr.br}

\author[label1]{André Grégio}
\ead{gregio@inf.ufpr.br}

\cortext[cor1]{Corresponding author.}
\address[label1]{Federal University of Paraná -- Cel. Francisco Heráclito dos Santos, 100 -- Curitiba, Paraná, Brazil}
\address[label2]{Victoria University of Wellington -- Cotton Building, Victoria University of Wellington, Kelburn, Wellington 6012, New Zealand}

\begin{abstract}
Malware is a major threat to computer systems and imposes many challenges to cyber security. Targeted threats, such as ransomware, cause millions of dollars in losses every year. 
The constant increase of malware infections has been motivating popular antiviruses (AVs) to develop dedicated detection strategies, which include meticulously crafted machine learning (ML) pipelines. However, malware developers unceasingly change their samples' features to bypass detection. 
\reviewthree{This constant evolution of malware samples causes changes to the data distribution (i.e., concept drifts) that directly affect ML model detection rates, something not considered in the majority of the literature work.}
\reviewthree{In this work, we evaluate the impact of concept drift on malware classifiers for two Android datasets: DREBIN ($\approx$130K apps) and a subset of AndroZoo ($\approx$285K apps). We used these datasets to train an Adaptive Random Forest (ARF) classifier, as well as a Stochastic Gradient Descent (SGD) classifier. We also ordered all datasets samples using their VirusTotal submission timestamp and then extracted features from their textual attributes using two algorithms (Word2Vec and TF-IDF). Then, we conducted experiments comparing both feature extractors, classifiers, as well as four drift detectors (Drift Detection Method, Early Drift Detection Method, ADaptive WINdowing, and Kolmogorov-Smirnov WINdowing) to determine the best approach for real environments. Finally, we compare some possible approaches to mitigate concept drift and propose a novel data stream pipeline that updates both the classifier and the feature extractor.}
\reviewthree{To do so,} we conducted a longitudinal evaluation by (i) classifying malware samples collected over nine years (2009-2018), (ii) reviewing concept drift detection algorithms to attest its pervasiveness, (iii) comparing distinct ML approaches to mitigate the issue, \reviewthree{and (iv) proposing an ML data stream pipeline that outperformed literature approaches, achieving an improvement of 22.05 percentage points of F1Score in the DREBIN dataset, and 8.77 in the AndroZoo dataset.}
\end{abstract}


\end{frontmatter}

\section{Introduction}
\label{introduction}
Countering malware is a major 
concern for most networked systems, since they can cause 
billions of dollars in loss~\cite{ransom}. 
The growth of malware
infections~\cite{infections} enables the development of multiple
detection strategies, including machine
learning (ML) classifiers tailored for malware, which have been adopted by 
the most popular AntiViruses (AVs)~\cite{Gandotra:2014,Kantchelian:2013:AAD:2517312.2517320,Jordaney2017}.
However, malware samples are very dynamic
pieces of code -- usually distributed over the Web~\cite{10.1145/2501654.2501663} and constantly evolving to survive -- quickly turning AVs into outdated mechanisms that present lower detection
rates over time. This phenomenon is 
known as concept 
drift~\cite{Gama:2014:SCD:2597757.2523813}, and requires AVs to periodically update 
their ML classifiers~\cite{nfs}. Malware concept drift is an understudied problem in the literature, with the few works that address it usually focusing on
achieving high accuracy rates 
for a temporally localized dataset, instead of aiming for long-term
detection due to malware evolution. 
Moreover, the community
considers ML a powerful ally for malware detection, given its ability to respond faster to new threats~\cite{Gibert2020}.

\reviewthree{Being the most used
operating system worldwide, Android has more than 2 billion
monthly active devices~\cite{pcworld_android},
with almost 40\% of prevalence in the
operating system market, surpassing
Microsoft Windows in
2018~\cite{statcounter}. As a widespread
platform, Android is more affected by malware
evolution and distribution, rendering its AVs
vulnerable to concept drift effects, which 
makes the need to adapt existing solutions.}

In this work, we evaluate the impact of concept drift  on malware classifiers for 
two Android datasets: DREBIN~\cite{arp2014}
($\approx$130K apps)
and a subset of AndroZoo~\cite{androzoo}
($\approx$285K apps).
%
For our longitudinal evaluation, we collected malware samples over
nine years (2009-2018) and used them to train an 
 Adaptive Random Forest (ARF) classifier~\cite{Gomes2017}, as well as 
\review{a Stochastic Gradient Descent (SGD) classifier~\cite{scikit-learn}}. \reviewthree{Our goal is
to answer the following questions:
(i) is concept drift a generalized phenomenon and not 
an issue of a particular dataset? (ii) is it important
to update the feature extractor (and not only the classifier)
when a concept drift is detected? (iii) how can we consider the feature 
extractor in the malware detection pipeline? (iv) which drift detector is the 
best for this task? (v) is concept drift somehow
related to changes in the Android ecosystem?}
To do so, we ordered all datasets samples 
using their VirusTotal~\cite{virustotal}
submission timestamp and then extracted
features from their textual attributes using
two algorithms (Word2Vec~\cite{w2v} and
TF-IDF~\cite{tfidf}). After that, we 
conducted experiments comparing both 
feature extractors, \review{classifiers,} as well as 
\review{four} drift detectors (Drift Detection
Method~\cite{Gama:2014:SCD:2597757.2523813}, 
Early Drift
Detection Method~\cite{baena2006}, 
ADaptive WINdowing~\cite{adwin}, \review{and
Kolmogorov-Smirnov WINdowing~\cite{Raab_2020}}) 
to determine the best approach for real environments.  
We also compare some possible 
approaches to mitigate 
concept drift 
and propose a novel method based on the data stream
pipeline that outperformed current solutions by updating both the classifier and the feature extractor. We 
highlight the need for also updating the feature extractor, given 
that it is as essential as the classifier
itself to achieve increased detection rates due to new
features appearing over time. 
Therefore, we 
propose a realistic data stream learning pipeline, including
the feature extractor in the loop.
\reviewthree{We also show with our experiments that concept drift is a generalized phenomenon in Android malware.}
Finally, we discuss the impact of changes 
on the Android ecosystem in our classifiers by 
comparing feature changes detected over time, and the implications of our findings. It is worth notice that all code and datasets used 
will be publicly  available\footnote{\url{https://www.kaggle.com/datasets/fabriciojoc/fast-furious-malware-data-stream}},
as well
as the implementation of our data stream learning 
pipeline using \texttt{scikit-multiflow}~\cite{skmultiflow}\footnote{\url{https://github.com/fabriciojoc/scikit-multiflow}}, 
an open-source ML framework for stream data.

This article is organized as follows: 
we compare our work with the literature
in Section~\ref{related_work};
we introduce our methodology in
Section~\ref{methodology}; 
we describe the machine learning 
background in 
Section~\ref{sec:feature_extraction};
we present the experiments results in
Section~\ref{experiments};
we discuss our findings
in Section~\ref{discussion},
and draw our conclusions
in Section~\ref{conclusion}.


\section{Related Work}
\label{related_work}

The literature on malware detection using machine learning is 
extensive~\cite{Gandotra:2014}.  
Usually, the primary concern 
of most works is to
achieve 100\% of accuracy using different 
representations and 
models, 
ignoring the fact that
malware samples evolve as time goes by. 
Few papers 
consider concept drift in this context, such as  
Masud et al., which were (at the best of our knowledge) the first to treat malware detection as a data stream
classification problem \reviewthree{like us, 
but} proposing an ensemble 
of classifiers trained from consecutive
chunks of data using $v$-fold partitioning of them and reducing 
classification error compared to other ensembles~\cite{Masud2008}.
They also presented a feature extraction and selection technique for
data streams that do not have any fixed feature set (\reviewthree{the opposite of our approach})
based on 
information gain.
Bach et al. combined two models: a stable one, based on all data, and 
a reactive one, based on a short window of recent data, to determine
when to replace the current stable model by computing the difference 
in their accuracy, assuming that the stable one performs worse than 
the reactive when the concept changes~\cite{Bach2008}, \reviewthree{very similar to what we do when the drift detector reaches a warning level.} 

Singh et al. proposed
two measures to track concept drift in static features of malware 
families \reviewthree{(which we do by using concept drift detectors, but in a malware detection task)}: relative temporal similarity (based on the 
similarity score 
between two time-ordered pairs of samples and are used to infer 
the direction of concept drift) and 
meta-features (based on summarization of information 
from a large number of features)~\cite{Singh2012}, claiming to provide paths to  
 further exploration of drift in malware detection models. 
Narayanan et al.
presented an online machine learning based 
framework, named DroidOL to handle concept drift and detect
malware~\cite{Narayanan2016} using
control-flow sub-graph features 
in an online 
classifier, 
which adapts to the 
malware drift 
by updating the model more aggressively 
when the error is large and less aggressively, otherwise, \reviewthree{which can also be done by drift detectors}. 
Deo et al.
proposed the use of Venn-Abers predictors to measure the 
quality of classification tasks and identify concept drift~\cite{Deo2016}. 

Jordaney et al. presented Transcend, a framework to identify concept
drift in classification models which compares the samples used to 
train the models with those seen during deployment,
computing algorithm credibility and
confidence to measure the quality of the produced results
and detect concept drift~\cite{Jordaney2017}. 
Anderson et al. have shown that, 
by using reinforcement learning to generate adversarial samples,
it is possible to retrain a model and make these attacks less
effective, also protecting it against possible drifts, 
given that this technique hardens a model against worst-case
inputs~\cite{anderson2018learning}, \reviewthree{something that can be improved even more by retraining the feature extractor as we do}. Pendlebury et al. reported
that some results are inflated by spatial bias, caused by the
wrong distribution of training and testing sets, and temporal
bias, caused by incorrect time splits into these same sets~\cite{tesseract}, \reviewthree{which is one of the reasons why we collected the timestamps in our datasets and used data streams}. 
They introduced an evaluation framework called 
\textsc{Tesseract} to compare malware classifiers in a
realistic setting and confirmed that earlier published results
are biased. \reviewthree{We reinforce that in our findings by comparing different experiments with a real-world simulation using data streams}. Ceschin et al. compared a set of experiments that 
use batch machine-learning models with ones that take 
into account the change of concept (data streams), emphasizing
the need to update the decision model immediately
after a concept drift occurs~\cite{nfs}. \reviewthree{In contrast, we do not update only the decision model, but also the feature extractor when drift occurs}. The authors also show that
the malware concept drift is related to their concept
evolution due to the appearance of new malware families.

\review{Mariconti et al. created \textsc{MaMaDroid}, an Android malware
detection system that uses a behavioral model, in the form of a Markov 
chain, to extract features from the API calls performed by an 
app~\cite{10.1145/3313391}. The solution proposed was tested with a 
dataset containing 44K samples, collected over six years, 
and presented a good detection rate and the ability to keep its 
performance for long periods (at least five years according to their
experiments).} \reviewthree{We increased the length of the observation window by using malware samples collected over nine years}. 
\review{Cai et al. compared their approach~\cite{cai2018preliminary, 10.1145/3183440.3195004},
which uses 52 selected metrics concerning sensitive data accesses via APIs
(using dynamic analysis) as features, with \textsc{MaMaDroid}. According 
to their experiments, their approach managed to remain stable for five years,
while \textsc{MaMaDroid} only kept the same performance for two years.
A very similar approach was compared with other four methods in literature
and all of them presented an overall f1score bellow 72\%~\cite{8802672}. }


Xu et al. proposed DroidEvolver, an Android malware detection
system that can be automatically updated without any human
involvement, requiring neither retraining nor true labels to 
update itself~\cite{Xu2019}. The authors use online learning 
techniques with evolving feature sets and pseudo labels, keeping
a pool of different detection models and calculating a juvenilization
indicator that determines when to update its feature
set and each detection model. \reviewthree{We compared our approach with DroidEvolver and showed that it outperformed it.}
Finally, Gibert et al. presented
 research 
challenges of 
state-of-the-art 
techniques for malware classification, exemplifying the concept drift as one of them~\cite{Gibert2020}.

\review{Zhang et al. designed \textsc{APIGraph}, a framework to detect evolved Android malware that groups similar API calls into clusters 
based on the Android API official documentation~\cite{10.1145/3372297.3417291}. 
Applying the aforementioned technique to other Android malware classifiers, the authors
reduced the labeling efforts when combined with \textsc{Tesseract}~\cite{tesseract}.}

\reviewthree{Finally, different from other approaches listed, Massimo Ficco presented an ensemble detector that exploits diversity in the detection algorithms by using both generic and specialized detectors (trained to detect certain malware types). The author also presents a mechanism that explores how the length of the observation time window can affect the detection accuracy and speed of different combinations of detectors during the detection~\cite{10.1109/TC.2021.3082002}.}


Our main contribution in this work is to
apply data stream based
machine learning algorithms to malware classification, 
proposing an important improvement in the pipeline that 
makes feature vectors correspond to the actual concept. The proposed solution, to the best of our knowledge, was not considered 
before and is as important as updating
the classifier itself. 
To test and validate our proposal, we use two datasets containing almost 415K android
apps, showing that it outperforms traditional 
data stream solutions. We also include an analysis of how certain features
change over time, correlating it with a cybersecurity background.


\section{Methodology} 
\label{methodology}

\subsection{Threat Model and Assumptions}

\reviewtwo{Our threat model considers an antivirus engine (AV) for the Android
platform since it is the market share leader. Thus, successful malware infections affect a large number of users. 
For scientific purposes, we considered an AV entirely based on Machine
Learning (ML) as implemented by many of the malware detectors cited
in Section~\ref{related_work}. In
practice, however, it does not imply that an AV must use only this
detection method: it can be complemented with any other detection
approach envisioned by the AV vendor. Our detection model is
completely static (features are retrieved directly from the APK
files) because static malware detection is the most popular and fastest way to
triage malware samples.
Similarly to the above discussion, the use of static detectors do not imply that an AV should not use
dynamic components, but that our research focuses on improving the
static detection component. In our proposed AV model, the APK files are
inspected as soon as they are placed in the Android filesystem, i.e., it does not rely on any other system information except
the APK files themselves. It is worth emphasizing that our goal is not to implement an actual AV but to highlight the need for updating ML models based on classifiers. Therefore, we 
simulated the behavior of an online AV using offline experiments that 
use data streams, simplifying our solution implementation. The
details of the simulated ML model are presented below.
}

\subsection{Data Stream}
\label{subsec:data_stream_cycle}

\begin{figure}[ht!] 
    \centering
    \includegraphics[width=.95\columnwidth]{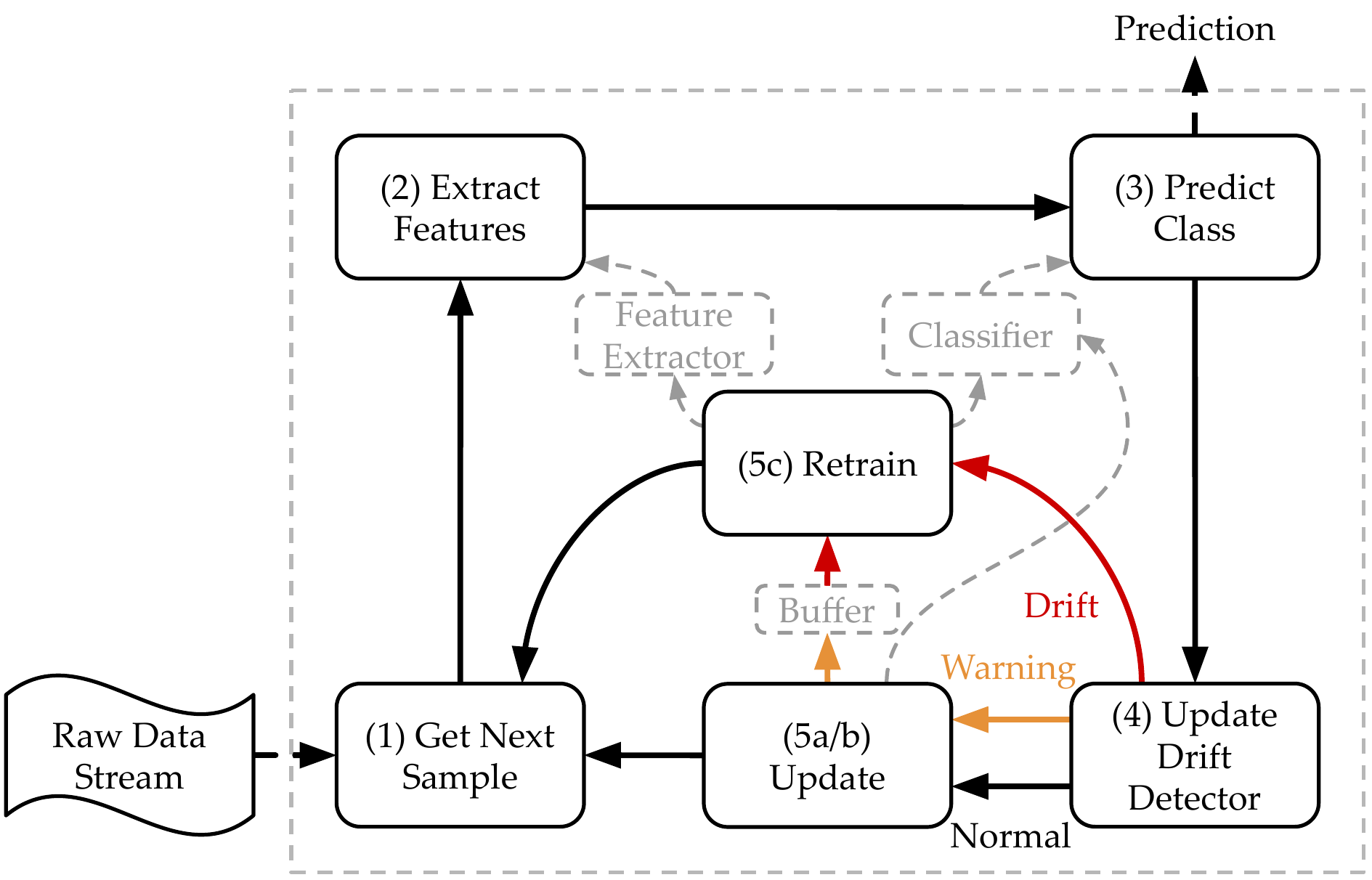}
    \caption{\textbf{Data Stream Pipeline.} Every time a new sample is obtained from the data stream, its features are extracted and presented to a classifier, generating a prediction, which is used by a drift detector that defines the next step: update the
    classifier or retrain both classifier and feature extractor.}
    \label{fig:datastream}
\end{figure}

Since our goal is to evaluate the occurrence of concept drift in malware classifiers, we analyzed malware detection evolution over time using a data stream abstraction for the input data. 
In a traditional data stream learning problem that includes concept drift, the classifier is updated with new samples when a change occurs---usually the ones that caused the drift~\cite{Gama:2014:SCD:2597757.2523813}. Our data stream pipeline also considers the feature extractor under changes, according to the following five steps shown in  Figure~\ref{fig:datastream}: 
\begin{enumerate}
    \item Obtain a new sample $X$ from the raw data stream;
    \item Extract features from $X$ using a feature extractor $E$, trained with previous data; 
    \item Predict the class of the new sample $X$ using the classifier $C$, trained with previous data; 
    \item With the prediction from $C$, update the drift detector $D$ to check the drift level (defined by authors~\cite{skmultiflow});
    \item According to the drift level, three paths can be followed, all of them making the pipeline restarts at \textbf{Step 1}:
    \begin{enumerate}[label=\alph*]
        \item \textbf{Normal:} incrementally \textbf{update} $C$ with $X$;
        \item \textbf{Warning:} incrementally \textbf{update} $C$ with $X$ and add $X$ to a buffer;
        \item \textbf{Drift:} \textbf{retrain} both $E$ and $C$ using only the data collected during the warning level (from the buffer build during this level), creating a new extractor and classifier.
    \end{enumerate}
\end{enumerate}
\subsection{Datasets}
A challenge to detect concept drift in malware
classifiers is to properly identify the sample's
date to allow temporal evaluations. Since malware
samples are collected in the wild and they may be spreading for some time, no actual creation
date is available. As an approximation for
it,
we considered each sample's
first appearance in VirusTotal~\cite{virustotal}, 
a website that
analyzes thousands of samples every day.
Malware samples were ordered by their
``first seen'' dates, which allows us to 
create a data stream representing a real-world 
scenario, where new 
samples are released daily, thus 
requiring
malware classifiers to be
updated~\cite{tesseract}. 

\begin{figure*}[t!]
    \centering
    \begin{subfigure}[t]{0.465\textwidth}
        \includegraphics[width=.95\columnwidth]{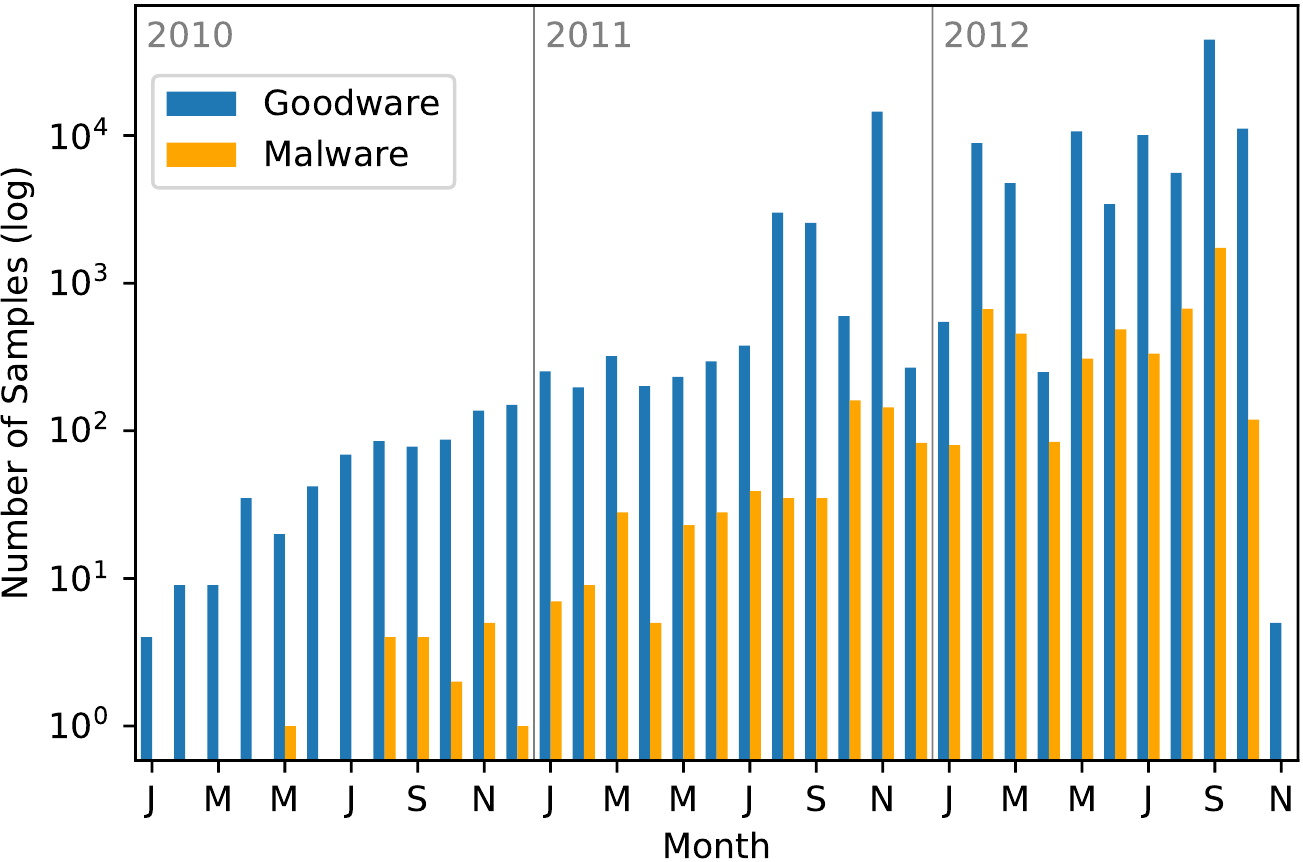}
        \caption{\textbf{DREBIN dataset distribution by month.} Very few samples are shown in 2009 (not shown), 2010, 2013 (not shown) and 2014 (not shown). The majority of them are in 2011 and 2012.} 
        \label{fig:distribution_drebin}
    \end{subfigure}
    \qquad
    \begin{subfigure}[t]{0.465\textwidth}
        \includegraphics[width=.95\columnwidth]{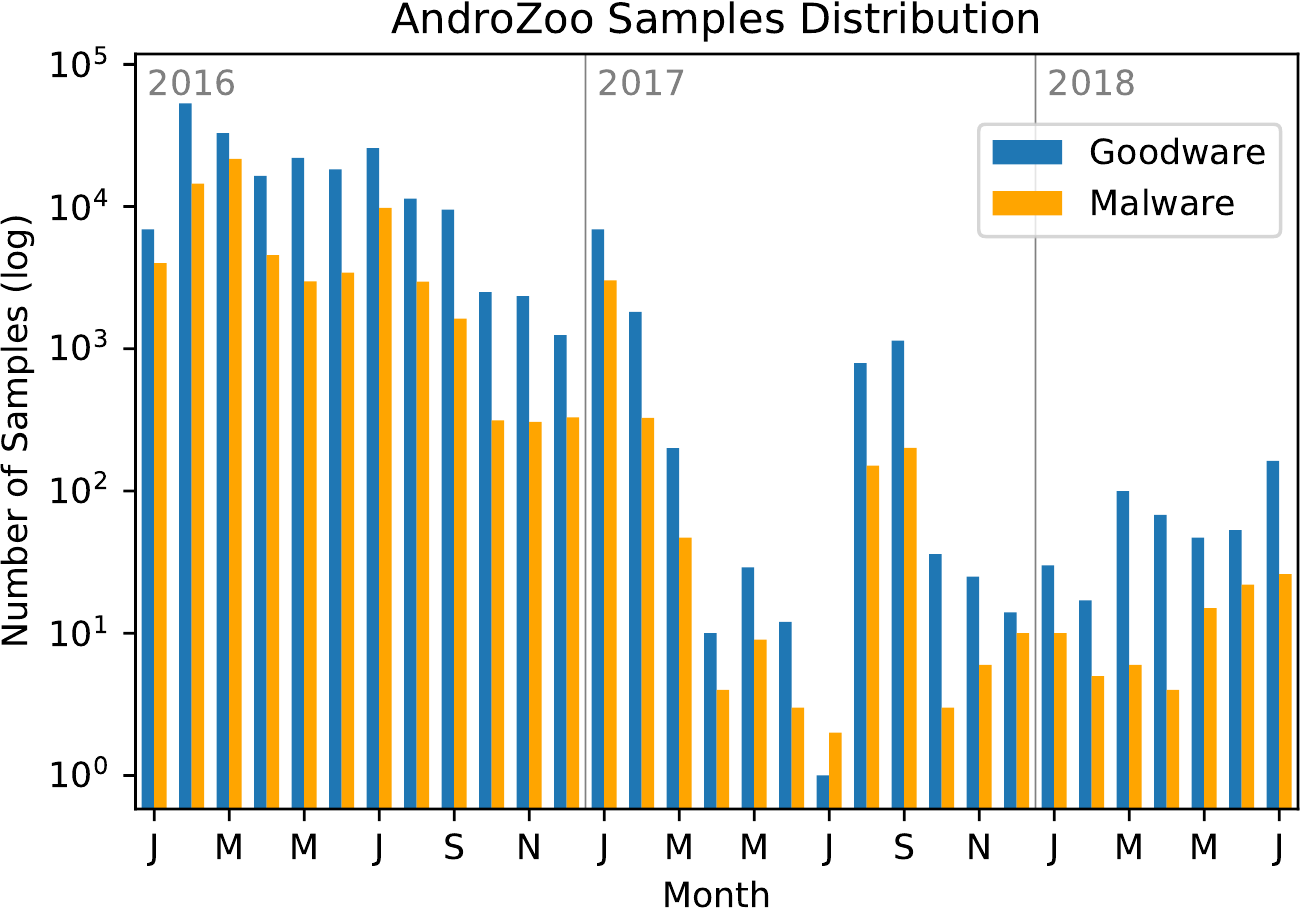}
        \caption{\textbf{AndroZoo dataset distribution by month.} The majority of samples are in 2016, with fewer malware showing in 2017. February 2016 and March 2016 are the months with more prevalence of goodware and malware, repectively.}
        \label{fig:distribution_androzoo}
    \end{subfigure}
    \caption{Datasets distribution over time.}
    \label{fig:distribution}
\end{figure*}

In our experiments, 
we considered
attributes vectors provided by the
authors of DREBIN~\cite{arp2014}, 
composed of ten textual attributes
(API calls, permissions, URLs, 
etc),
which are publicly available to download and contain
$123,453$ benign 
and $5,560$
malicious Android applications. We show DREBIN's distribution in Figure~\ref{fig:distribution_drebin}.
We also considered a subset of Android applications reports
provided by AndroZoo API~\cite{androzoo}, 
composed
of eight textual attributes (resources names,
source code classes and methods, manifest permissions 
 etc.) and contains $213,928$ benign and $70,340$ malicious applications. We show the distribution of our AndroZoo subset in Figure~\ref{fig:distribution_androzoo}:
it keeps the same goodware and malware distribution as the original 
dataset, which originally is composed by most of 10 million apps.

It is important to notice that we   
are using the attributes that were already 
extracted statically from 
them~\cite{arp2014,androzoo}, since we do not have access to applications 
binaries, packages, or source codes. In addition, the 
 timing information provided by the 
authors of the DREBIN~\cite{arp2014} 
and VirusTotal differs. According to 
Arp et al.~\cite{arp2014},
their samples were collected from August 2010 
to October 2012. However, our version is based
on VirusTotal 
appearance date, which showed
us that very few samples were already analyzed by
VirusTotal (48 in 2009) and some of them were just analyzed
after the establishment of the collection (37 in 2013 and 2014), probably following
the dataset publicly release. 
We do not show these
samples in Figure~\ref{fig:distribution} 
for better visualization.

\begin{figure*}[ht!]
    \centering
    \centering
    \begin{subfigure}[b]{.95\textwidth}
        \includegraphics[width=\linewidth]{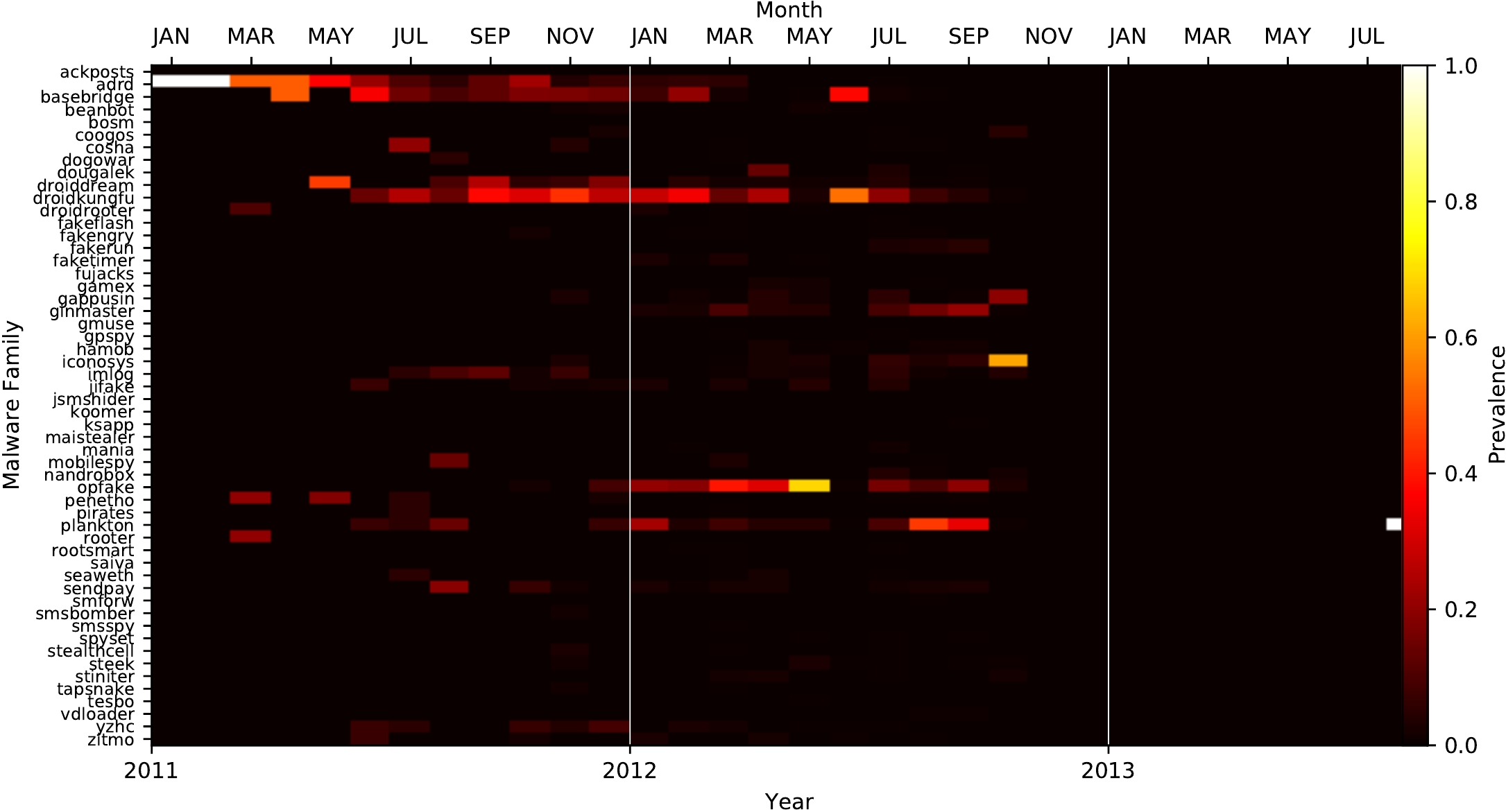}
        \caption{DREBIN dataset.}
        \label{fig:mw_family_drebin}
    \end{subfigure}
    ~
    \centering
    \centering
    \begin{subfigure}[b]{.95\textwidth}
        \includegraphics[width=\linewidth]{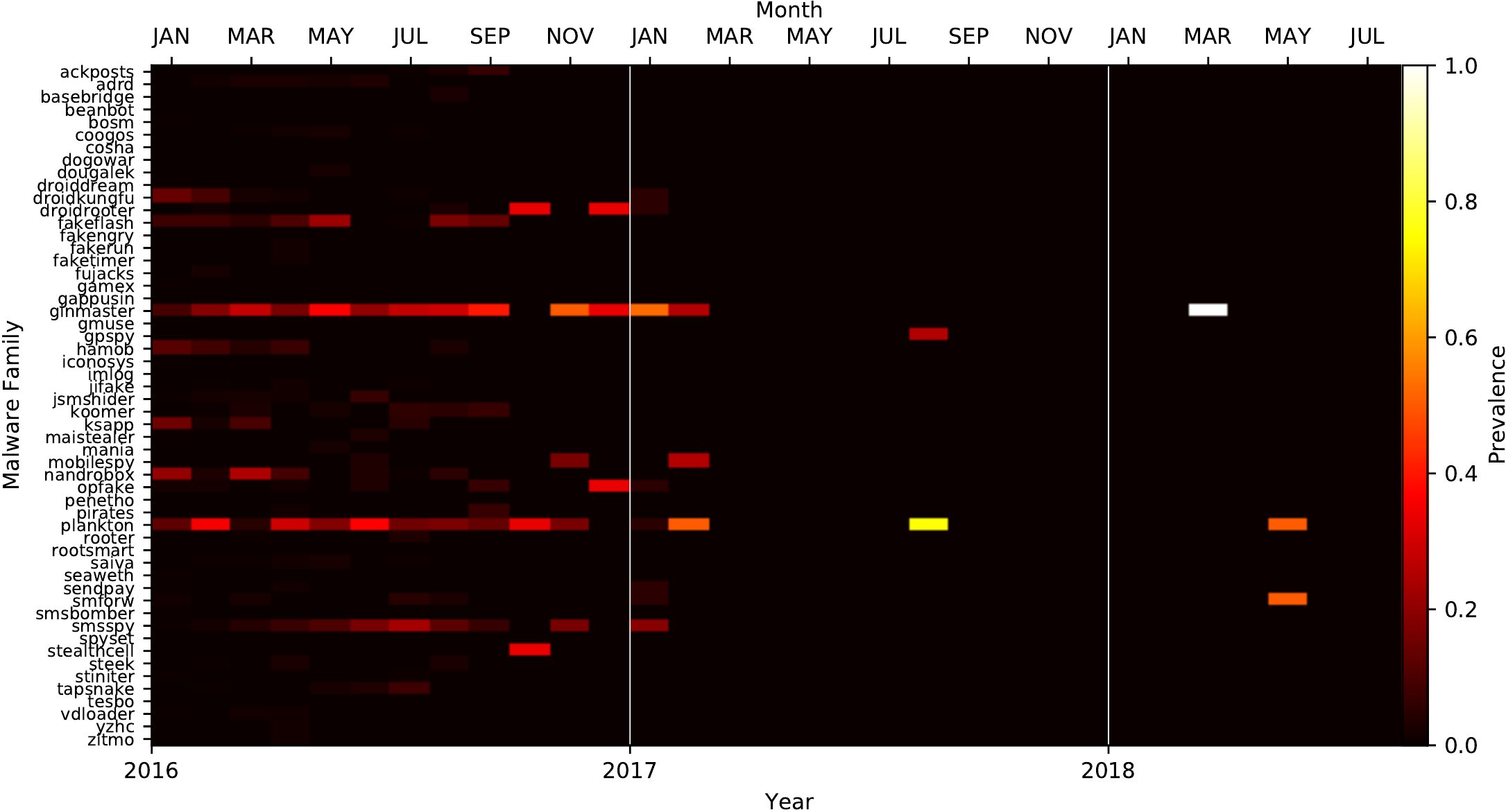}
        \caption{AndroZoo dataset.}
        \label{fig:mw_family_androzoo}
    \end{subfigure}
    \caption{\textbf{Malware family distribution.} Intersection of families from both datasets shows 
    evidences of class evolution, given that they are very different.} 
    \label{fig:mw_family}
\end{figure*}

Furthermore, both datasets reflect two characteristics
of the real world that challenge the 
development of efficient ML malware detectors:
(i) long-term variations (2009-2014, for DREBIN and
2016-2018 for AndroZoo);
and (ii) 
class imbalance. For example, in DREBIN, 
whereas more than
40K goodware
were collected in Sep/2009,
only 1,750 malware samples were collected
in the same period. Whereas class
imbalance is out of this work's scope,
we considered both datasets as suitable
for our evaluations due to the long-term
characteristic, which challenges ML 
classifiers to learn multiple concepts
over time. 
Finally, to evidence the difference of both datasets, 
we created heat maps containing the 
prevalence of a subset of malware 
families created using the intersection
of families from them (54 families), as shown in Figures~\ref{fig:mw_family_drebin} and~\ref{fig:mw_family_androzoo}. 


    


\section{Machine Learning Algorithms}
\label{sec:feature_extraction}


\subsection{Representation}
%
A typical way
to represent malware for ML is to create a
vector using the sample textual attributes,
such as API calls, permissions, URLs, 
providers, intents activities, service 
receivers, and others.
%
In our work, we represented them using Word2Vec~\cite{w2v}
and TF-IDF~\cite{tfidf}, since they are
widely used representations for text
classification. Besides, both use the
training data to generate the representation
of all samples, which allows us to test our 
hypothesis that the drift affects not only 
the classifier but also the feature 
extractor. 
\subsection{Concept Drift Detectors}
%
%
We used \review{four} concept drift detection algorithms in our work: 
DDM (Drift Detection Method~\cite{Gama:2014:SCD:2597757.2523813}), 
EDDM
(Early Drift Detection Method~\cite{baena2006}),
ADWIN (ADaptive WINdowing~\cite{adwin}), and
\review{KSWIN (Kolmogorov-Smirnov WINdowing)~\cite{Raab_2020}}.
Both DDM and EDDM are online supervised methods based on 
sequential error 
monitoring, i.e., each 
incoming example is processed separately estimating the
sequential error rate. Therefore, they assume that 
the increase of consecutive error rate suggests the
occurrence of concept drift. DDM directly uses the error
rate, while EDDM uses the distance-error rate, which 
measures the number of examples between two 
errors~\cite{baena2006}.
These errors trigger two 
levels: warning and drift. The warning level suggests 
that the concept starts to drift, then an alternative
classifier is updated using the samples which rely on 
this level. The drift level suggests that the concept
drift occurred, and the alternative classifier build 
during the warning level replaces the current classifier.
ADWIN keeps statistics from sliding windows of variable 
size, 
used to compute the average of the change
observed by cutting them in different points. 
If the difference between two windows is greater
than a pre-defined threshold, a concept drift is detected, 
and the data from the first window is discarded~\cite{adwin}.
\review{KSWIN uses a sliding window of fixed size to compare
the most recent samples of the window with the remaining 
ones by using Kolmogorov-Smirnov (KS) statistical 
test~\cite{Raab_2020}.}
Different from the other two methods, ADWIN and KSWIN has no
warning level, which made it necessary to adapt our 
data stream ML pipeline.
For instance, when a change occurs \review{in ADWIN,} out of the window data is discarded and the remaining samples are 
used to retrain the both the classifier and feature extractor.
\review{Finally, when a change occurs in KSWIN, only the 
most recent samples of the window are kept and used
to retrain the classifier and feature extractor.}

\subsection{Classifiers}
In all evaluations, we developed classification
models leveraging Word2Vec and TF-IDF
representations and normalized using a MinMax 
technique~\cite{scikit-learn}. For all cases,
we used \review{as classifiers} Adaptive Random Forest~\cite{Gomes2017} 
without its internal drift detectors (working as 
a Random Forest for data streams), since its widespread use in
the malware detection literature and
has the best overall performance~\cite{nfs},
\review{and  Stochastic Gradient Descent (SGD) 
classifier~\cite{scikit-learn}, which is one
of the fastest online classifiers in 
scikit-learn~\cite{scikit-learn}.}
Both the classifier and drift detectors 
were configured using the same
hyper-parameters proposed by 
the authors~\cite{Gomes2017,skmultiflow}.

\section{Experiments}
\label{experiments}

\subsection{The Best-Case Scenario
for AVs (\textit{ML Cross-Validation})}
\label{exp1}
In the first 
experiment, we classify all samples
together to compare which feature extraction
algorithm is the best and report baseline
results. We tested several parameters for
both algorithms and fixed the vocabulary 
size in $100$ for TF-IDF (top-100 features
ordered by term frequency) and created
projections with $100$ dimensions for Word2Vec,
resulting in $1,000$ and $800$ features for each app in
both cases, for DREBIN and AndroZoo, respectively.
All results are reported after 10-fold
cross-validation procedures, 
a method commonly used in ML to evaluate
models because its results are less prone 
to biases (note that we are training new
classifiers and feature extractors
at every iteration of the cross-validation process). 
In practice, folding the dataset
implies that the AV company has a mixed 
view of both past and future threats,
despite temporal effects, which is the
best scenario for AV operation and ML
evaluation.



\begin{table*}[t!]
\centering
\resizebox{\textwidth}{!}{%
\begin{tabular}{@{}cccccccccc@{}}
\toprule
\multirow{2}{*}{\textbf{Classifier}}                                     & \multirow{2}{*}{\textbf{Algorithm}} & \multicolumn{4}{c}{\textbf{DREBIN Dataset}}                                  & \multicolumn{4}{c}{\textbf{AndroZoo Dataset}}                                \\
                                                                         &                                     & \textbf{Accuracy} & \textbf{F1Score} & \textbf{Recall}  & \textbf{Precision} & \textbf{Accuracy} & \textbf{F1Score} & \textbf{Recall}  & \textbf{Precision} \\ \midrule
\multirow{2}{*}{\begin{tabular}[c]{@{}c@{}}Random\\ Forest\end{tabular}} & Word2Vec                            & 99.09\%           & 88.73\%          & 82.12\%          & \textbf{96.48\%}   & 90.52\%           & 76.27\%          & 66.03\%          & 90.29\%            \\
                                                                         & TF-IDF                              & \textbf{99.23\%}  & \textbf{90.63\%} & \textbf{85.85\%} & 96.31\%            & \textbf{91.54\%}  & \textbf{79.30\%} & \textbf{70.25\%} & \textbf{91.03\%}   \\ \midrule
\multirow{2}{*}{SGD}                                                     & Word2Vec                            & 98.29\%           & 78.28\%          & 70.90\%          & 87.36\%            & 85.41\%           & 60.05\%          & 47.52\%          & 81.54\%            \\
                                                                         & TF-IDF                              & \textbf{98.63\%}  & \textbf{83.26\%} & \textbf{78.49\%} & \textbf{88.66\%}   & \textbf{88.74\%}  & \textbf{71.57\%} & \textbf{61.43\%} & \textbf{85.73\%}   \\ \bottomrule
\end{tabular}%
}
\caption{\textbf{Cross-Validation.} Mixing past and future threats is the best scenario for AVs in both datasets.}
\label{table:cross-validation}
\end{table*}



Table~\ref{table:cross-validation}
presents the results obtained in this
experiment for both DREBIN and AndroZoo 
datasets \review{using Adaptive Random Forest (ARF)
and Stochastic Gradient Descent (SGD)}, 
highlighting the performance 
of TF-IDF, which was better than Word2Vec 
in all metrics, except in precision when
classifying the DREBIN dataset. 
It means that Word2Vec is slightly better 
to detect goodware (particularly in DREBIN 
dataset) since its precision 
is higher (i.e., fewer FPs) and TF-IDF 
is better to detect malware due to its 
higher recall (i.e., less FNs). In general,
we conclude that TF-IDF is better than
Word2Vec since its accuracy and f1score 
are higher. Moreover, we notice that its 
f1score is not as high as the accuracy, 
which is near 100\%, indicating that one 
of the classes (malware) is more difficult 
to predict than the other (goodware).
Regardless of small differences, we observe
that ML classifiers perform significantly 
well 
when samples of all periods are mixed, 
even in a more complex dataset such as
AndroZoo, since they can learn
features from all periods. 

\subsection{On Classification Failure
(\textit{Temporal Classification})}
\label{exp2}
Although currently used classification methodology helps reducing
dataset biases, it would demand knowledge about future threats to work properly.
AV companies
train their classifiers using data from
past samples and leverage them to predict
future threats, expecting to present 
the same characteristics as past ones. 
However, malware samples are 
very dynamic, thus this strategy is the
worst-case scenario for AV companies. 
To demonstrate the effects of predicting
future threats based on past data, we split
our datasets in two: we used the first half (oldest samples) to train  our classifiers, which were then used to predict the newest samples from the second half.
The results of Table~\ref{table:half_experiment}
indicate a drop in all metrics when compared to the 10-fold experiment
in both DREBIN and AndroZoo datasets \review{and also sugest the occurrence of concept drift on malware samples, given that the recall is much smaller than the previous experiment.}

                   

\begin{table*}[t!] 
\centering
\caption{\textbf{Temporal Evaluation.} Predicting future threats based on data from
the past is the worst-case for AVs.}
\label{table:half_experiment}
\resizebox{\textwidth}{!}{%
\begin{tabular}{@{}cccccccccc@{}}
\toprule
\multirow{2}{*}{\textbf{Classifier}}                                     & \multirow{2}{*}{\textbf{Algorithm}} & \multicolumn{4}{c}{\textbf{DREBIN Dataset}}                                  & \multicolumn{4}{c}{\textbf{AndroZoo Dataset}}                                \\
                                                                         &                                     & \textbf{Accuracy} & \textbf{F1Score} & \textbf{Recall}  & \textbf{Precision} & \textbf{Accuracy} & \textbf{F1Score} & \textbf{Recall}  & \textbf{Precision} \\ \midrule
\multirow{2}{*}{\begin{tabular}[c]{@{}c@{}}Random\\ Forest\end{tabular}} & Word2Vec                            & 97.66\%           & 62.58\%          & 46.31\%          & 96.47\%            & 87.55\%           & 53.95\%          & 38.71\%          & 88.96\%            \\
                                                                         & TF-IDF                              & \textbf{98.20\%}  & \textbf{73.26\%} & \textbf{58.36\%} & \textbf{98.39\%}   & \textbf{88.20\%}  & \textbf{57.13\%} & \textbf{41.71\%} & \textbf{90.63\%}   \\ \midrule
\multirow{2}{*}{SGD}                                                     & Word2Vec                            & 97.52\%           & 65.14\%          & 55.08\%          & 79.70\%            & 85.81\%           & 47.04\%          & 33.44\%          & 79.28\%            \\
                                                                         & TF-IDF                              & \textbf{98.15\%}  & \textbf{75.18\%} & \textbf{66.42\%} & \textbf{86.61\%}   & \textbf{86.96\%}  & \textbf{52.79\%} & \textbf{38.68\%} & \textbf{83.07\%}   \\ \bottomrule

\end{tabular}%
}
\end{table*}


Due to dataset imbalance in both datasets, we notice a bias toward goodware detection (very small accuracy decrease) and significant qualitative differences for f1score and recall (much worse than before).
The precision score was very similar to the
cross-validation experiment, and better
in the case of TF-IDF when classifying
DREBIN, showing that goodware samples
remain similar in the ``future'' while
malware samples evolved. 
Word2Vec and TF-IDF lost, \review{in average}, 
\review{about 16} percentage points of their
f1score in DREBIN and \review{about 19} percentage points
for this same metric in AndroZoo. \review{In addition, the model's recall rates
significantly dropped in both Word2Vec and TF-IDF when classifying
DREBIN and AndroZoo, regardless of the models used.} This indicates that detecting 
unforeseen malware only from a single set of past
data is a hard task. These results
highlight the need of developing better
continuous learning approaches for effective
malware detection.

\subsection{Real-World Scenario
(\textit{Windowed Classifier})}
\label{exp3}

Since 
static classifiers
are a bad strategy, AV
companies adopt continuous updating
procedures as samples 
are collected and identified.
From a ML perspective, they adopt
an incremental stream learning 
method~\cite{pinage2016}, which we 
call Incremental Windowed Classifier (IWC).
\reviewtwo{Notice that this is the same approach used by
other authors in the literature, but they
propose the use of distinct attributes and features only, instead of a new stream pipeline~\cite{10.1145/3372297.3417291, 10.1145/3371924}.}
To evaluate
the impact of this approach, we
divided the datasets into two groups, one containing samples released until a
certain month of a year for training,
and the other with samples released 
one month after that said month for
testing. For example, considering
Jan/2012, the training set contained
samples that were created until Jan/2012
(cumulative) and the validation set
contained samples created only in 
Feb/2012 (a month later). We tested
\review{both Adaptive Random Forest (ARF)~\cite{Gomes2017}
and Stochastic Gradient Descent (SGD) classifier~\cite{scikit-learn}}
with TF-IDF (100 features for each textual attribute),
given its better performance in previous
experiments, and excluded months with no samples.
Every month, both classifier and feature 
extractor were retrained, generating new
classifiers and feature extractors.

\begin{figure*}[t!] 
    \centering
    \begin{subfigure}[b]{0.435\textwidth}
        \includegraphics[width=\columnwidth]{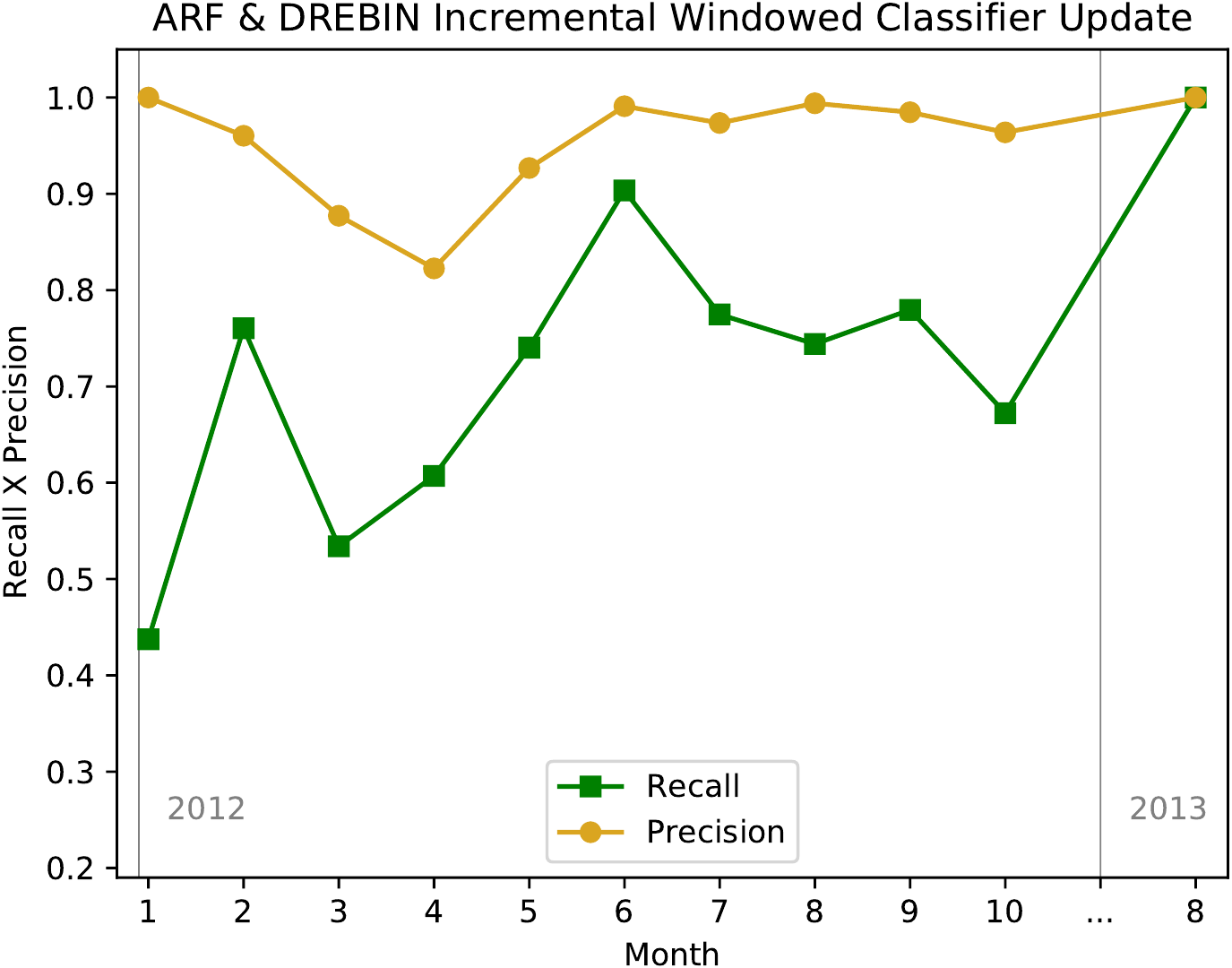}
        \caption{Adaptive Random Forest and DREBIN dataset.}
        \label{fig:exp3_drebin}
    \end{subfigure}
    \qquad
    \begin{subfigure}[b]{0.48\textwidth}
        \centering
        \includegraphics[width=.89\columnwidth]{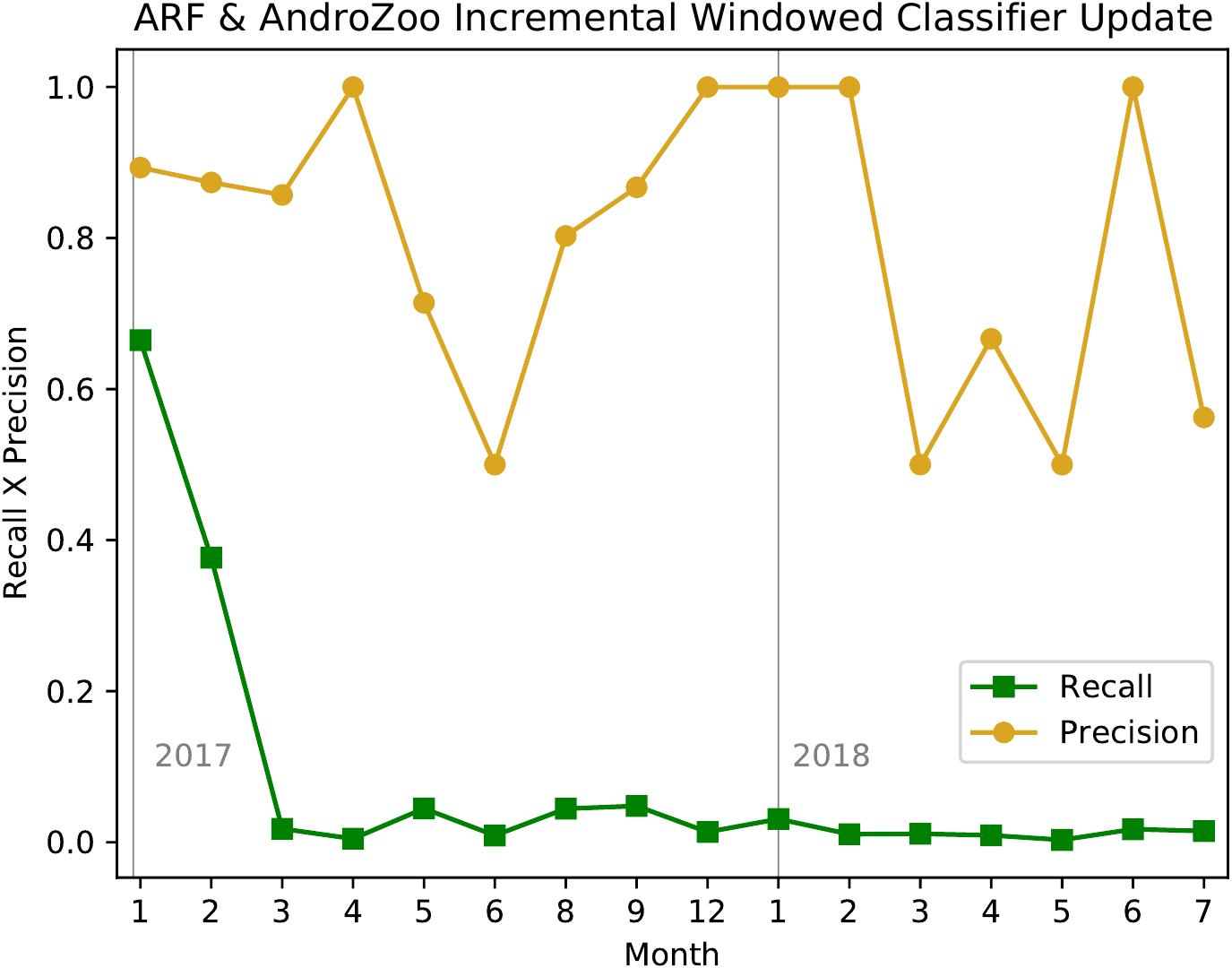}
        \caption{Adaptive Random Forest and AndroZoo dataset.}
        \label{fig:exp3_androzoo}
    \end{subfigure}
    
    \begin{subfigure}[b]{0.435\textwidth}
        \includegraphics[width=\columnwidth]{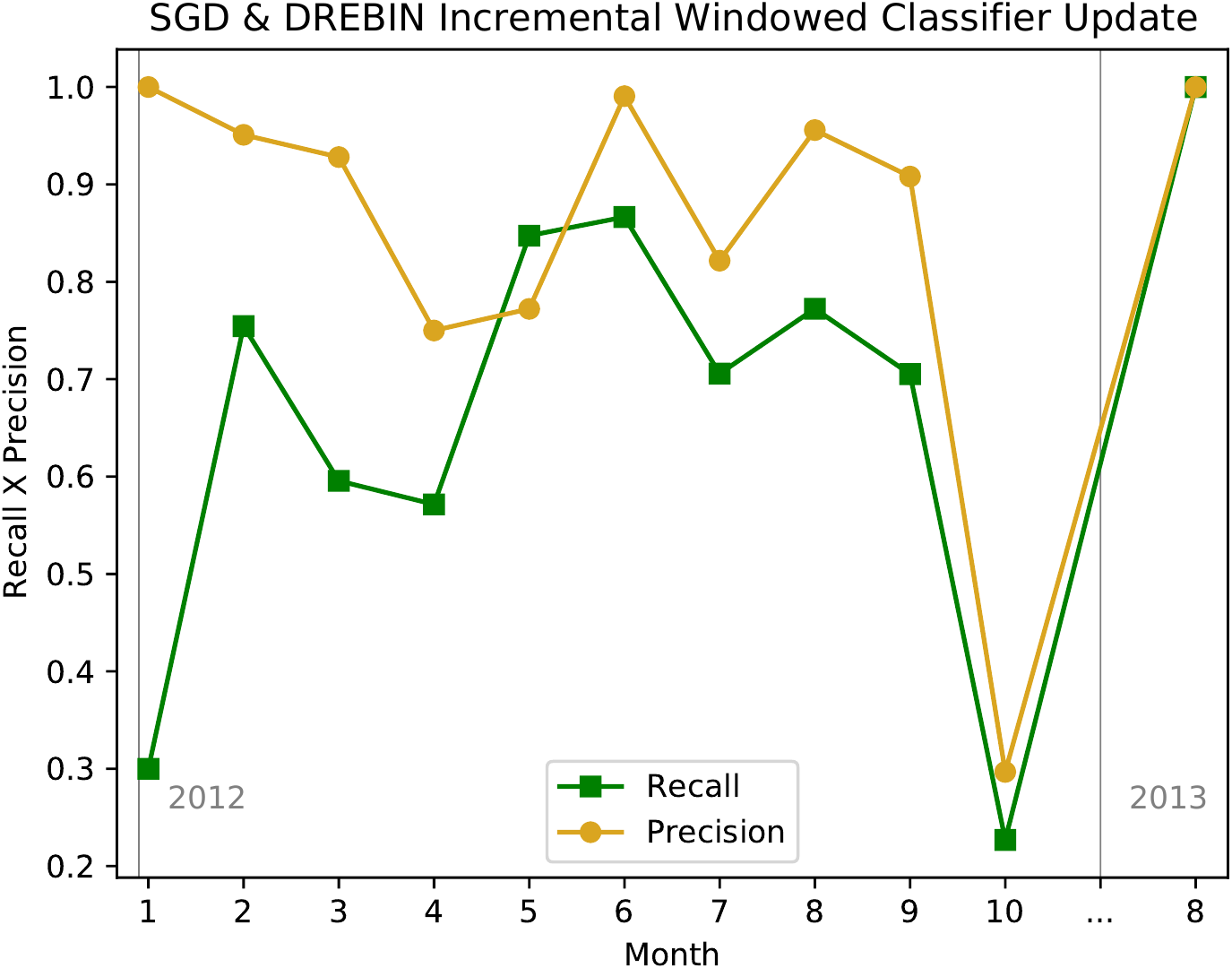}
        \caption{\review{SGD and DREBIN dataset.}}
        \label{fig:exp3_drebin_sgd}
    \end{subfigure}
    \qquad
    \begin{subfigure}[b]{0.48\textwidth}
        \centering
        \includegraphics[width=.89\columnwidth]{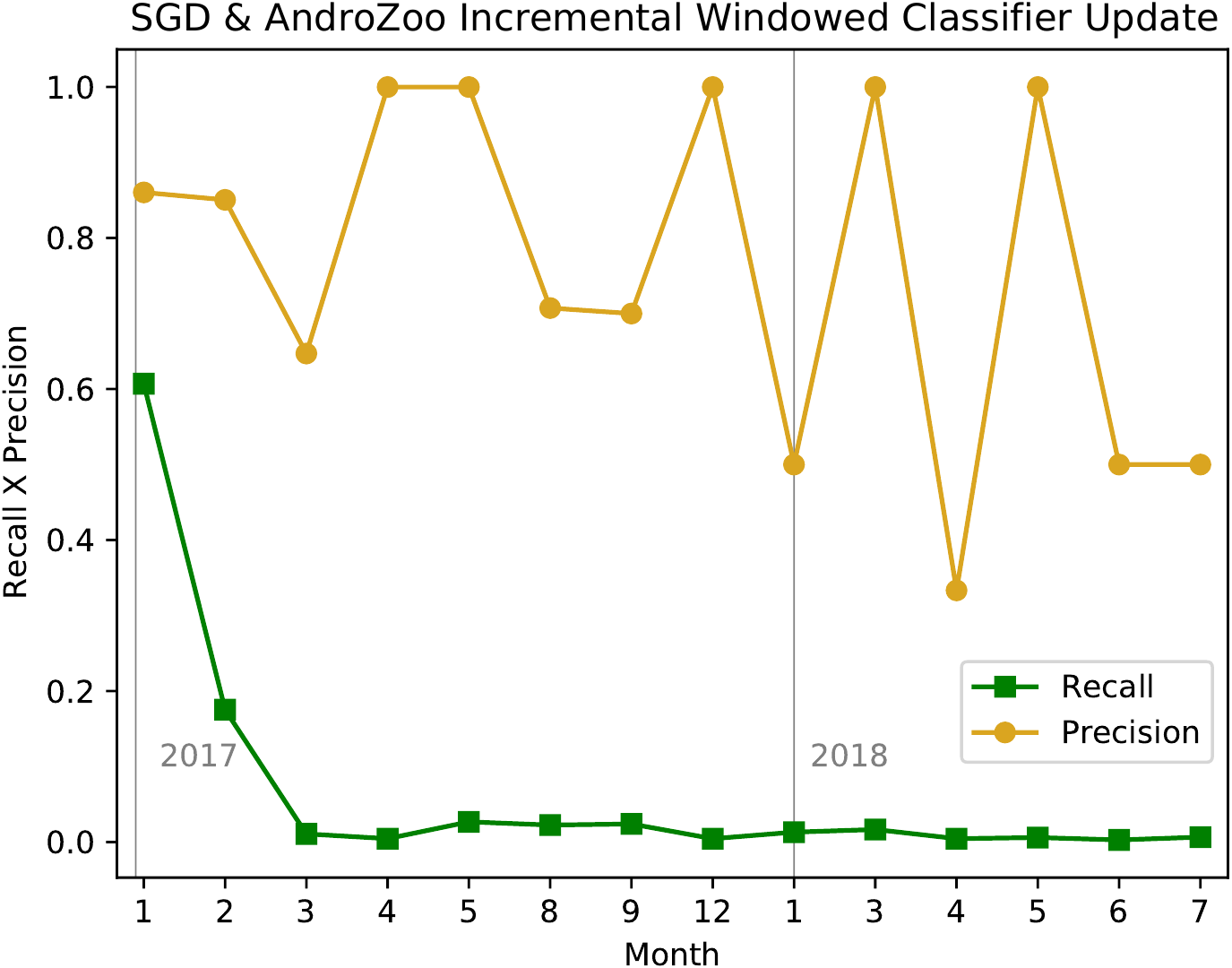}
        \caption{\review{SGD and AndroZoo dataset.}}
        \label{fig:exp3_androzoo_sgd}
    \end{subfigure}
    \caption{\textbf{Continuous Learning.} Recall and precision
    while incrementally retraining both classifier \review{(Adaptive Random Forest and Stochastic Gradient Descent classifier)} and feature extractor.}
    \label{fig:exp3}
\end{figure*}

The results for the DREBIN dataset shown in
\review{Figures~\ref{fig:exp3_drebin} and~\ref{fig:exp3_drebin_sgd}}
indicate a \review{drop in both precision and recall} rates.
\review{For Adaptive Random Forest}, precision
drops to about 85\% in April/2012 and 
increases in the following months to almost 100\% in 
August/2013. The Recall starts with the worst
result (about 40\%), in January/2012 and 
reaches almost 100\% in August/2013.
The average recall was 70.3\% and average
precision, 95.36\%. 
\review{For SGD, the worst precision
is reported in October/2012 (about 30\%), increasing to almost 100\% in
August/2013. The same happen with the recall, which drops to almost 20\%
in October/2012 and increases to almost 100\% in August/2013. 
Finally, the average recall was 66.78\% and average
precision, 85.22\%.}
On the one hand,
the growth of precision and recall
in some periods indicate 
that continuously updating
the classifier might increase classification
performance in comparison to using a single
classifier tested with samples from a past
period because the classifier learns
new patterns for the existing concepts 
(features) that may have been introduced 
over time. On the other hand, the drop in 
the precision and recall rates in some 
periods indicate the presence of 
new concepts (e.g., new malware features) 
that were not fully handled by the
classifiers.
When looking at the AndroZoo dataset results,
as shown in 
\review{Figures~\ref{fig:exp3_androzoo} and~\ref{fig:exp3_androzoo_sgd}},
it is possible to note a drastic fall in
recall as time goes by \review{in both cases}, 
dropping from \review{almost 70\% (ARF) and 60\% (SGD)} in January/2017 
to less than 5\% in March/2019, not exceeding \review{10\%} by the end of 
the stream. This indicates that AndroZoo is a much more
complex dataset with very different malware in distinct
periods, given that the recall did not behave as
this same experiment using the DREBIN dataset. 
In contrast, the precision remained
unstable.
\review{For Adaptive Random Forest,} it reaches 100\% in 
some periods, such as May/2017, and drops to almost 50\%, 
in June/2017, with quite similar behavior from the DREBIN 
dataset.
\review{For SGD, it also reaches 100\% in some periods, such as
April/2017, and drops to almost 30\% in April/2018.}
These results indicate that more 
than continuous retraining, AV companies 
need to develop classifiers fully able to
learn new concepts.

\subsection{Concept Drift Detection using Data Stream Pipeline (Fast \& Furious -- F\&F)}
\label{exp4}
Although the previous results indicate that
continuously learning from the past
is the best strategy for detecting
forthcoming threats, concept drift remains
challenging, even more when considering
results from the AndroZoo dataset. Therefore, we
present an evaluation of a drift detector
approach that could be deployed by AV
companies to automatically update their
classifier models. In our experiments, 
\review{we used both Adaptive Random Forest~\cite{Gomes2017}
and Stochastic Gradient Descent
as the classifiers}, DDM, EDDM, ADWIN, \review{and KSWIN} as drift 
detectors~\cite{skmultiflow}
(
with the same parameters as the 
authors), and TF-IDF (using 100 features for each textual
attribute) as a feature extractor. In the DREBIN 
dataset, we initialized the base classifier with data from
the first year, given that in the first months we have
just a few malware showing up. In the AndroZoo dataset,
we initialized it with data from the first month. We tested 
all drift detection algorithms in two 
circumstances when creating a new classifier, 
according to our data stream pipeline: 
(i) \textbf{Update:}
we just update the classifier with new 
samples (collected in warning level or
ADWIN window), which
reflects the fastest approach for
AV companies to react to new threats; 
(ii) \textbf{Retrain:} we extract all 
features from the raw data (also 
collected in warning level or ADWIN window) 
and all models are built from scratch 
again, i.e., both
classifier and feature extractor are 
retrained and a new vocabulary is built based 
on the words in the training set for each
textual attribute (more complete approach,
but time-consuming for AV companies).
To compare our solution with another
similar in the literature, we implemented 
our version of DroidEvolver~\cite{Xu2019}, 
replicating their approach using the same
feature representation as ours. It is important
to report here that we tried to use the authors'
source code, but they were not working properly
due to dependencies that could not be installed. 
Thus, we implemented an approximation of their
method by analyzing their paper and code, using
$\tau_{0} = 0.3$ and $\tau_{1} = 0.7$ with three
compatible online learning methods from 
scikit-learn~\cite{scikit-learn}: \texttt{SGDClassifier},
\texttt{PassiveAggressiveClassifier}, 
and \texttt{Perceptron}. Finally, during the execution of 
DroidEvolver, we noticed that it was detecting at least one
drift in one of its classifiers each iteration, making it
necessary to create a new parameter that checks for drift 
in an interval of \textit{steps} (iterations), in our case 
$500$ steps was selected according to our analysis.

\begin{figure*}[t!] 
    \centering
    \begin{subfigure}[b]{0.465\textwidth}
        \includegraphics[width=\columnwidth]{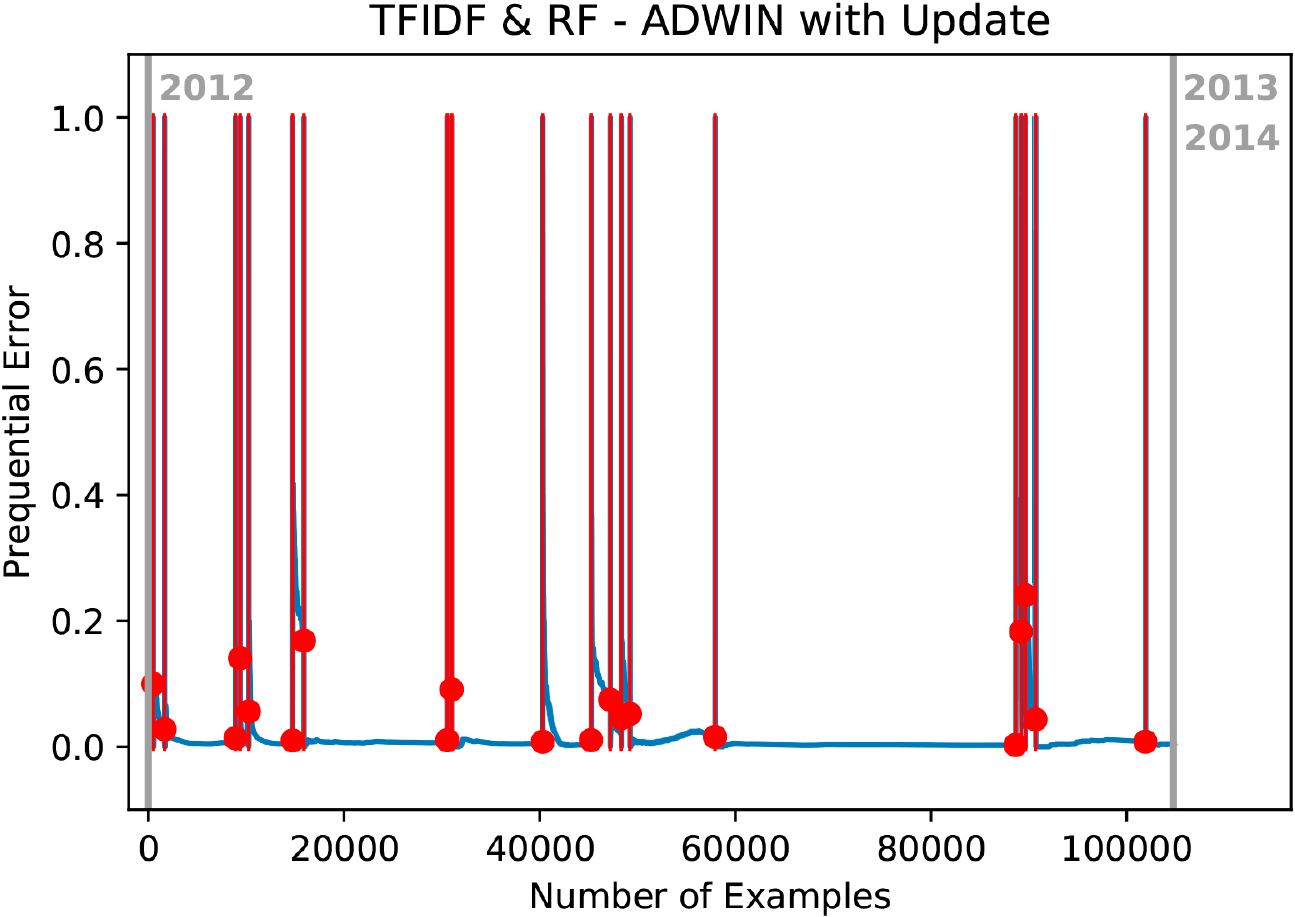}
        \caption{\reviewtwo{Adaptive Random Forest with ADWIN and Update.}}
        \label{fig:exp_4_adwin_update}
    \end{subfigure}
    \qquad
    \begin{subfigure}[b]{0.465\textwidth}
        \centering
        \includegraphics[width=\columnwidth]{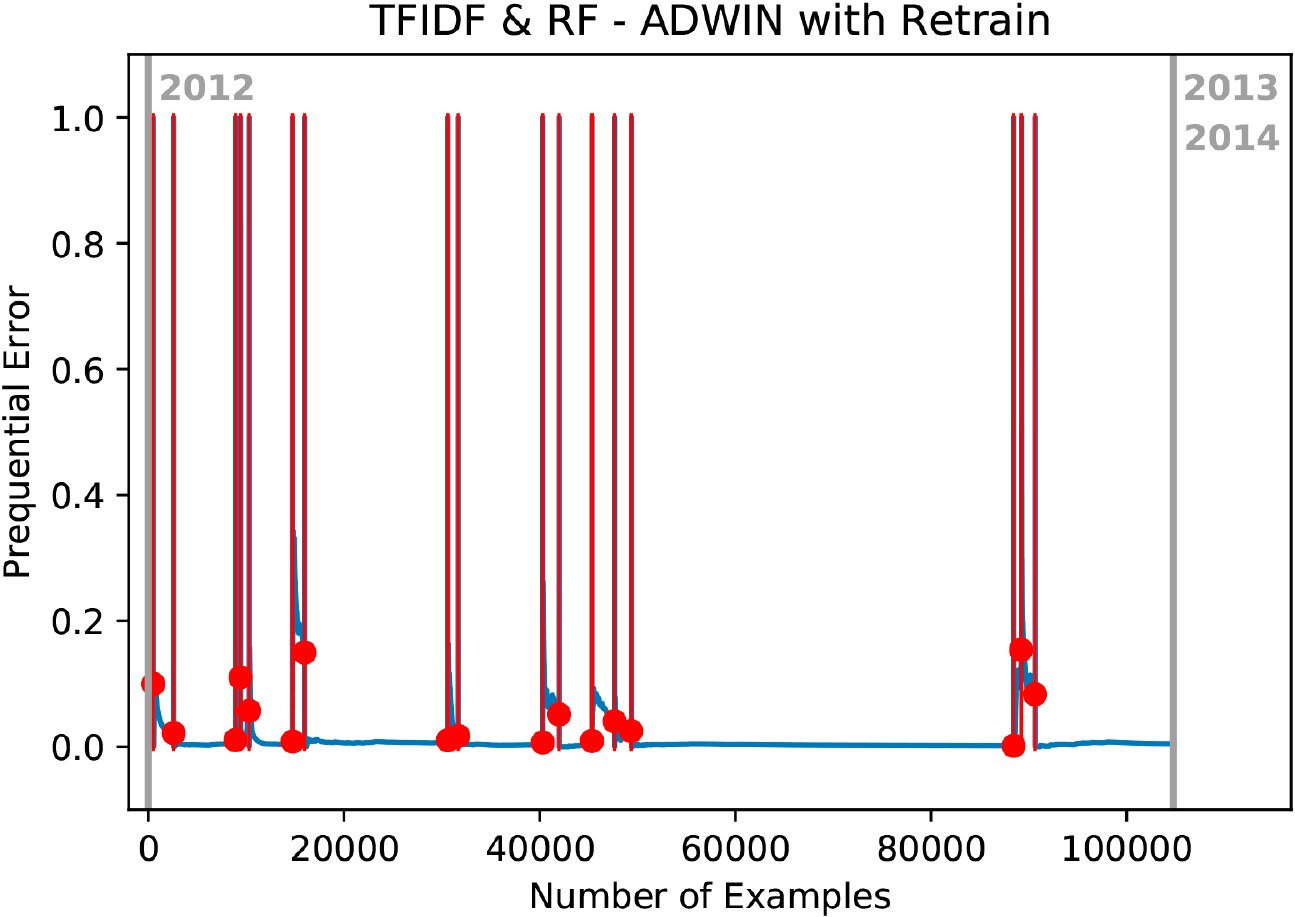}
        \caption{\reviewtwo{Adaptive Random Forest with ADWIN and Retrain.}}
        \label{fig:exp_4_adwin_retrain}
    \end{subfigure}
    \caption{\reviewtwo{\textbf{F\&F (U)pdate and (R)etrain prequential error and drift points as time goes by when using Adaptive Random Forest.} Despite being similar in some points, fewer drift points are detected when retraining the feature extractor, reducing the prequential error and increasing classification performance.}}
    \label{fig:exp_4_adwin}
\end{figure*}

Table~\ref{table:results} presents 
the results for DroidEvolver and our methods 
for both datasets. 
For DREBIN, \review{when using Adaptive Random Forest},
EDDM outperforms DDM methods'
classification performance in all scenarios,
\review{which was the opposite when using SGD classifier}.
\review{For both classifiers,} ADWIN outperforms EDDM, DDM, \review{and KSWIN}, providing the best overall performance
for retraining as well as for learning new features.
However, \review{only when using Adaptive Random Forest,} ADWIN with the update 
is better for precision, making it slightly better in reducing false 
positives, \review{and KSWIN is better for recall, making it better 
in reducing false negatives.} 
Moreover, retraining both the feature extractor and classifier makes it detect fewer drift points than the updating approach \review{(18 vs. 20 when using ARF, 13 vs 14 when using SGD)}. \review{Overall, SGD outperformed Adaptive Random Forest when classifying DREBIN, presenting an improvement in f1score of almost 16 percentage points.}
In comparison to DroidEvolver, our approach outperformed it in all metrics, with a relatively large margin. \reviewtwo{In Figure~\ref{fig:exp_4_adwin}, it is possible to observe the prequential error of Adaptive Random Forest with ADWIN using Update (Figure~\ref{fig:exp_4_adwin_update}) and Retrain (Figure~\ref{fig:exp_4_adwin_retrain}) strategies. Despite the Retrain strategy is similar to the update approach in some points, it has less drift points, a lower prequential error and, consequently, a better classification performance.}

When classifying AndroZoo, DDM outperforms EDDM, which was not a good drift detector for this specific dataset. ADWIN with retraining was the best method again, detecting 50 and 33 drift points versus 78 and 30 when using the update approach \review{with Adaptive Random Forest and SGD classifier, respectively},  outperforming all other methods, including DroidEvolver again. Finally, these results suggest that AV companies should keep investigating specific samples to increase overall detection. 

\begin{table*}[t!] 
\centering
\caption{\textbf{Overall results.} 
Considering IWC, DroidEvolver~\cite{Xu2019}, F\&F (U)pdate and (R)etrain strategies with multiple drift detectors and classifiers (Random Forest and SGD).}
\label{table:results}
\resizebox{\textwidth}{!}{%
\begin{tabular}{@{}cccccccccccc@{}}
\toprule
\multirow{2}{*}{\textbf{Classifier}}                                                         & \multirow{2}{*}{\textbf{Method}}                                   & \multicolumn{5}{c}{\textbf{DREBIN Dataset}}                                                    & \multicolumn{5}{c}{\textbf{AndroZoo Dataset}}                                                  \\
                                                                                             &                                                                    & \textbf{Accuracy} & \textbf{F1Score} & \textbf{Recall}  & \textbf{Precision} & \textbf{Drifts} & \textbf{Accuracy} & \textbf{F1Score} & \textbf{Recall}  & \textbf{Precision} & \textbf{Drifts} \\ \midrule
\textbf{\begin{tabular}[c]{@{}c@{}}Model\\Pool\end{tabular}}                                                                          & \textbf{\begin{tabular}[c]{@{}c@{}}DroidEvolver\\ \cite{Xu2019}\end{tabular}} & 97.27\%           & 67.14\%          & 59.14\%          & 77.64\%            & 69              & 87.09\%           & 66.28\%          & 56.17\%          & 80.83\%            & 22              \\ \midrule
\multirow{9}{*}{\textbf{\begin{tabular}[c]{@{}c@{}}Adaptive\\ Random\\ Forest\end{tabular}}} & \textbf{IWC}                                                       & 96.8\%            & 80\%             & 70.3\%           & 95.36\%            & N/A             & 82.99\%           & 11.4\%           & 8.25\%           & 79.61\%            & N/A             \\
                                                                                             & \textbf{DDM (U)}                                                   & 98.3\%            & 79.19\%          & 68.38\%          & 94.04\%            & 8               & 88.19\%           & 70.84\%          & 63.5\%           & 80.11\%            & 14              \\
                                                                                             & \textbf{DDM (R)}                                                   & 98.4\%            & 80.54\%          & 70.27\%          & 94.32\%            & 9               & 87.96\%           & 69.92\%          & 61.94\%          & 80.27\%            & 24              \\
                                                                                             & \textbf{EDDM (U)}                                                  & 98.53\%           & 82.27\%          & 71.84\%          & 96.26\%            & 27              & 78.28\%           & 39.52\%          & 37.23\%          & 42.11\%            & 118             \\
                                                                                             & \textbf{EDDM (R)}                                                  & 98.57\%           & 82.85\%          & 73.09\%          & 95.6\%             & 14              & 77.91\%           & 39.31\%          & 37.52\%          & 41.28\%            & 17              \\
                                                                                             & \textbf{ADWIN (U)}                                                 & 98.58\%           & 82.66\%          & 71.35\%          & \textbf{98.21\%}   & 20              & 86.41\%           & 65.86\%          & 58.02\%          & 76.15\%            & 78              \\
                                                                                             & \textbf{ADWIN (R)}                                                 & \textbf{98.71\%}  & \textbf{84.44\%} & 74.17\%          & 98.02\%            & 18              & 89.6\%            & \textbf{75.05\%} & \textbf{69.23\%} & 81.93\%            & 50              \\
                                                                                             & \textbf{KSWIN (U)}                                                 & 98.38\%           & 80.82\%          & 72.19\%          & 91.80\%            & 10              & 89.22\%           & 71.78\%          & 60.66\%          & 87.88\%            & 70              \\
                                                                                             & \textbf{KSWIN (R)}                                                 & 98.55\%           & 83.01\%          & \textbf{74.96\%} & 93.00\%            & 9               & \textbf{89.66\%}  & 73.22\%          & 62.56\%          & \textbf{88.26\%}   & 50              \\ \midrule
\multirow{9}{*}{\textbf{SGD}}                                                                & \textbf{IWC}                                                       & 96.33\%           & 73.40\%          & 66.78\%          & 85.22\%            & N/A             & 82.63\%           & 9.16\%           & 6.61\%           & 75.71\%            & N/A             \\
                                                                                             & \textbf{DDM (U)}                                                   & 98.84\%           & 87.56\%          & 86.29\%          & 88.88\%            & 3               & 88.05\%           & 71.17\%          & 65.30\%          & 78.20\%            & 7               \\
                                                                                             & \textbf{DDM (R)}                                                   & 98.85\%           & 87.67\%          & 86.69\%          & 88.66\%            & 3               & 89.10\%           & 73.81\%          & 67.98\%          & 80.73\%            & 4               \\
                                                                                             & \textbf{EDDM (U)}                                                  & 98.85\%           & 87.63\%          & 86.43\%          & 88.87\%            & 11              & 82.99\%           & 42.50\%          & 32.97\%          & 59.75\%            & 85              \\
                                                                                             & \textbf{EDDM (R)}                                                  & 98.78\%           & 87.05\%          & 86.61\%          & 87.49\%            & 20              & 82.77\%           & 41.80\%          & 32.45\%          & 58.72\%            & 73              \\
                                                                                             & \textbf{ADWIN (U)}                                                 & 98.95\%           & 88.68\%          & 87.03\%          & 90.38\%            & 14              & 89.39\%           & 73.98\%          & 66.74\%          & 82.97\%            & 30              \\
                                                                                             & \textbf{ADWIN (R)}                                                 & \textbf{99.00\%}  & \textbf{89.19\%} & \textbf{87.38\%} & \textbf{91.08\%}   & 13              & \textbf{89.49\%}  & \textbf{74.31\%} & 67.25\%          & \textbf{83.01\%}   & 33              \\
                                                                                             & \textbf{KSWIN (U)}                                                 & 98.93\%           & 88.45\%          & 86.83\%          & 90.13\%            & 1               & 88.23\%           & 72.41\%          & \textbf{68.34\%} & 77.00\%            & 48              \\
                                                                                             & \textbf{KSWIN (R)}                                                 & 98.85\%           & 87.71\%          & 87.28\%          & 88.15\%            & 2               & 85.34\%           & 67.05\%          & 65.99\%          & 68.14\%            & 70              \\ \bottomrule
\end{tabular}%
}
\end{table*}

\subsection{\reviewtwo{Multiple Time Spans}}

\reviewtwo{In the previous experiment, we showed that the use of drift detectors with the Retrain strategy improved  the classification performance. Although the experiment simulates a real-world stream, there might be potential biases from the use of the very same training and test sets. By proposing a new experiment, we mitigated the risk of biasing evaluations, since we have tested our solutions under different circumstances by using multiple time spans and reporting the average result.}

\reviewtwo{To evaluate the effectiveness of our approach in multiple conditions,
we applied our data stream pipeline to different training and test sets of our dataset: we splitted the two datasets (sorted by the samples' timestamps) into eleven folds (thus 10 consecutive epochs), with every fold containing the same amount of data, similar to a \textit{k}-fold cross-validation scheme. However, instead of using a single set for training as done in each iteration of \textit{k}-fold cross-validation, we increment the training set \textit{i} with the fold \textit{i+1}, and remove it from the test set at every iteration. This way, we create a cumulative training set and simulate the same scenario as the previous experiment, but starting the stream in different parts. In the end, we produced ten distinct results that present the effectiveness of our method under varied conditions, which worked as a \textit{k}-fold cross-validation. However, we focused on collecting the F1Score of each iteration).}

\begin{figure*}[t!] 
    \centering
    \begin{subfigure}[b]{0.465\textwidth}
        \includegraphics[width=\columnwidth]{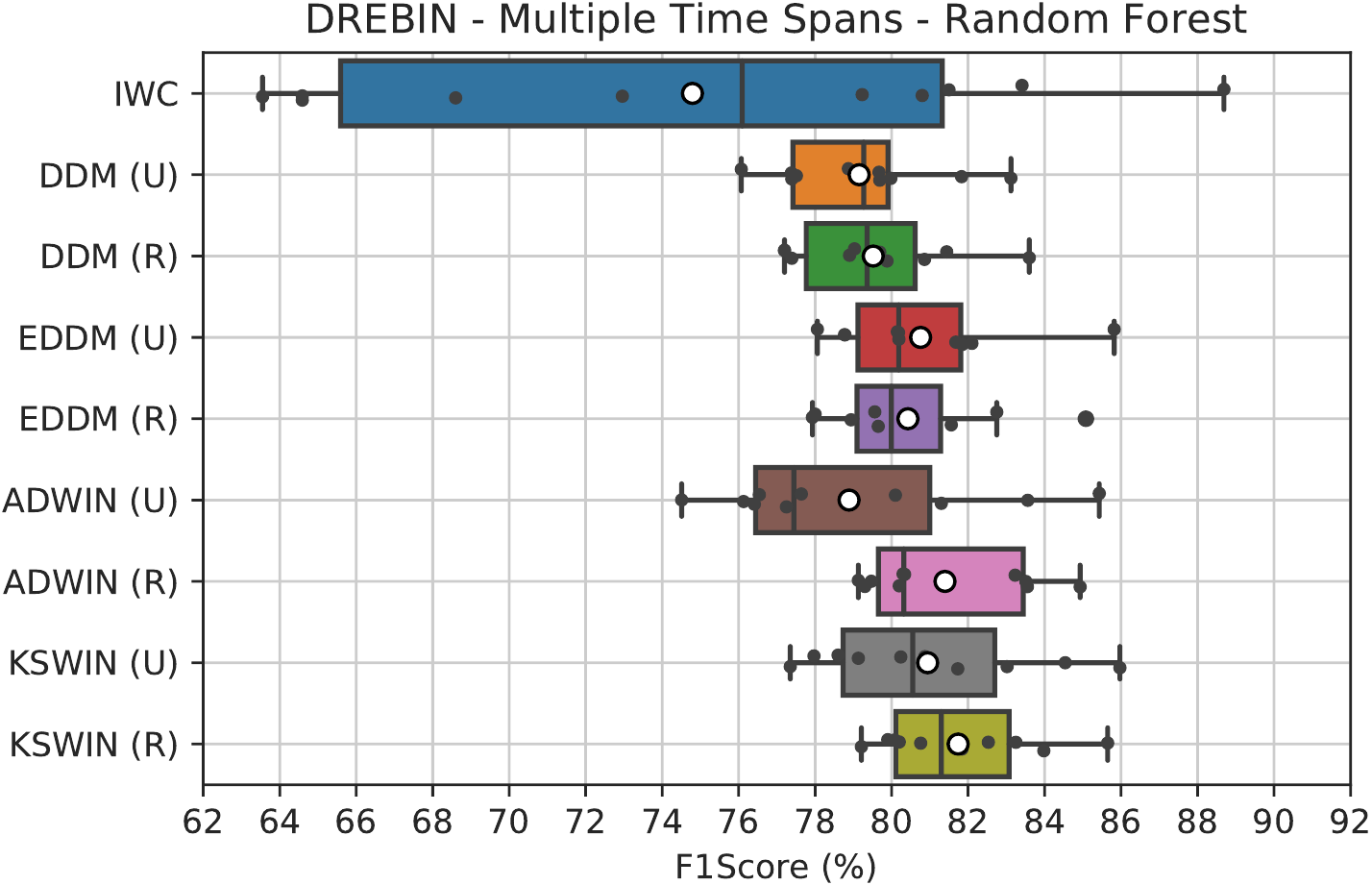}
        \caption{\reviewtwo{Adaptive Random Forest and DREBIN dataset.}}
        \label{fig:mts_drebin_rf}
    \end{subfigure}
    \qquad
    \begin{subfigure}[b]{0.465\textwidth}
        \centering
        \includegraphics[width=\columnwidth]{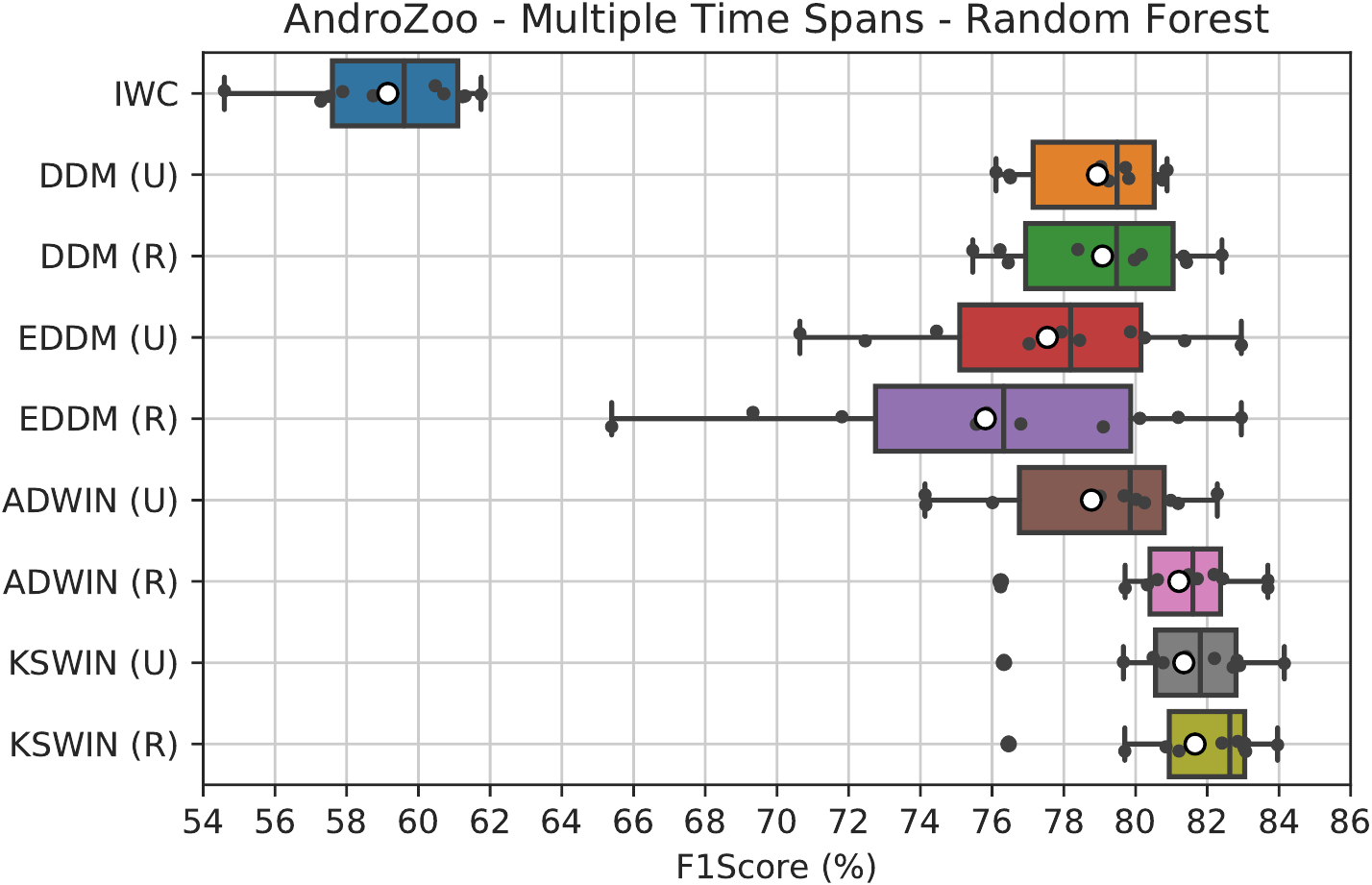}
        \caption{\reviewtwo{Adaptive Random Forest and AndroZoo dataset.}}
        \label{fig:mts_androzoo_rf}
    \end{subfigure}
    
    \begin{subfigure}[b]{0.465\textwidth}
        \includegraphics[width=\columnwidth]{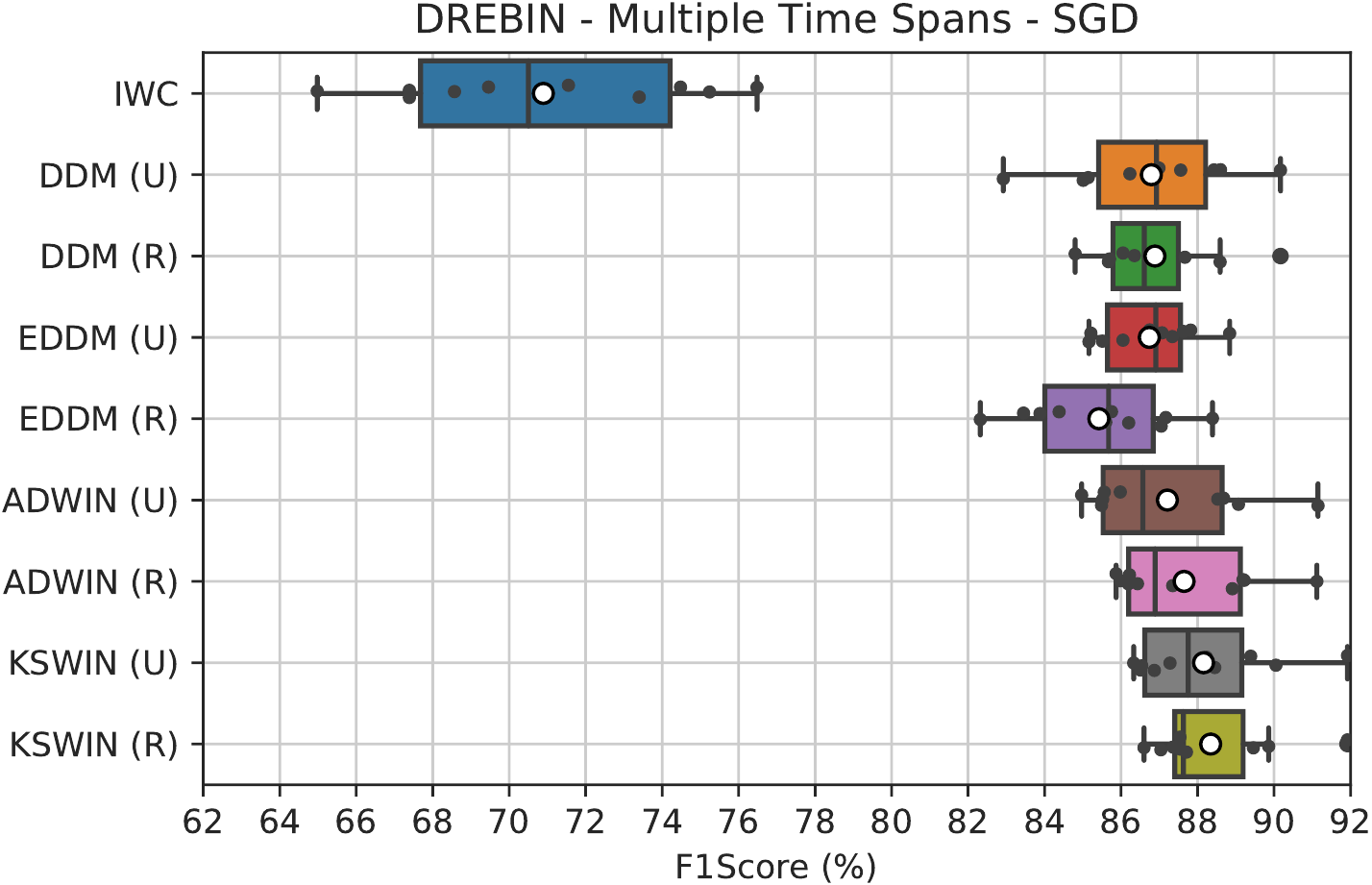}
        \caption{\reviewtwo{SGD and DREBIN dataset.}}
        \label{fig:mts_drebin_sgd}
    \end{subfigure}
    \qquad
    \begin{subfigure}[b]{0.465\textwidth}
        \centering
        \includegraphics[width=\columnwidth]{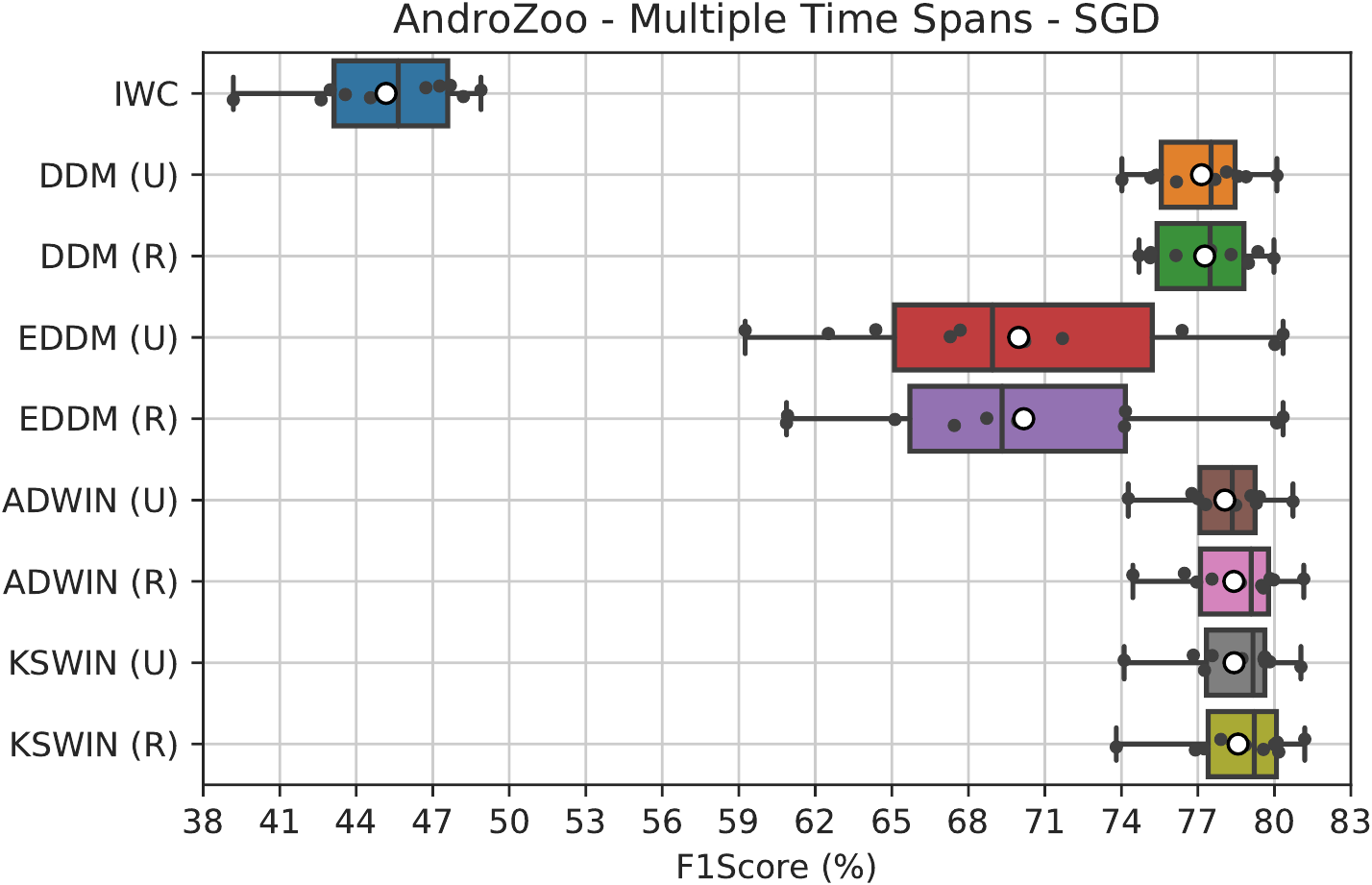}
        \caption{\reviewtwo{SGD and AndroZoo dataset.}}
        \label{fig:mts_androzoo_sgd}
    \end{subfigure}
    \caption{\textbf{\reviewtwo{Multiple Time Spans.}} \reviewtwo{F1Score distribution when experimenting with ten different sets of training and test samples. Black dots represent the distribution of all F1Scores, whereas white dots represent their average for each applied method.}}
    \label{fig:mts}
\end{figure*}

\reviewtwo{In the current experiment, we used all the methods presented in the previous section: both classifiers (Adaptive Random Forest and SGD), all drift detectors (DDM, EDDM, ADWIN, and KSWIN), and both methods ((U)pdate and (R)etrain). To accomplish a better presentation of the results, we chose a boxplot visualization containing the distribution of all F1Scores (black dots) and their average for each method (white dots), as shown in Figure~\ref{fig:mts}.}

\reviewtwo{In Figures~\ref{fig:mts_drebin_rf} and~\ref{fig:mts_drebin_sgd}, we present the results for the DREBIN dataset using Adaptive Random Forest and SGD as classifiers, respectively. 
The IWC method performed much better than the others with Random Forest; in the case of using ten folds to predict the last one, it reaches a better performance than the other methods, which indicates that the last fold may not be affected by concept drift. However, when we look at the other methods, KSWIN with retrain performed best, achieving a higher average F1Score and a low standard deviation. Also, as we saw in the previous section, SGD performed better than the Adaptive Random Forest applied to this dataset.}

\reviewtwo{In Figures~\ref{fig:mts_androzoo_rf} and~\ref{fig:mts_androzoo_sgd}, we  present the results for the AndroZoo dataset using Adaptive Random Forest and SGD as classifiers, respectively. In this scenario, IWC performance is much worse than any other method, which we believe is evidenced due to the complexity of this dataset (almost 285K samples). Again, in both classifiers, KSWIN with retrain method performed best, achieving the higher average F1Score, despite Adaptive Random Forest presenting better overall results and being almost 4 percentage points higher than SGD.}

\reviewtwo{Finally, the analysis of the results allows us to observe that (i) the more data we have in the data stream, the most difficult it becomes for IWC to keep a good performance; (ii) using KSWIN drift detector with retrain is the most recommended method for static android malware detection data streams; and (iii) we ensure that our results do not have potential biases. }


\subsection{Understanding Malware Evolution}
After we confirmed that concept drift is
prevalent in these malware datasets, we delved
into evolution details to understand the
reasons behind it and the lessons that mining
a malware dataset might teach us. To do so, 
we analyzed vocabulary changes over time
for both datasets and correlated our findings
of them. 

\begin{figure*}[t!] 
    \centering
    \begin{subfigure}[b]{.95\textwidth}
        \includegraphics[width=\linewidth]{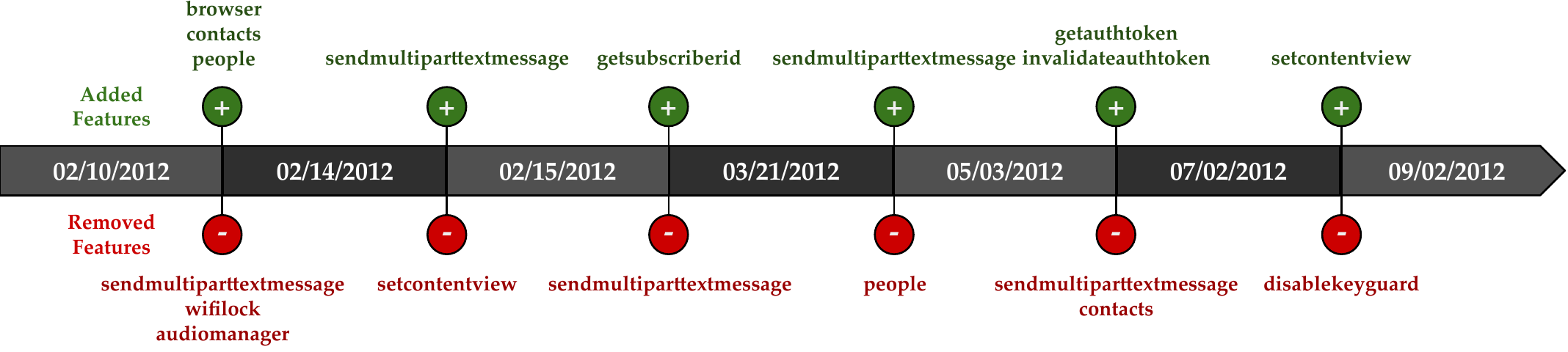}
        \caption{Vocabulary change timeline for API calls using DREBIN dataset.}
        \label{fig:drebin_vocabulary_change}
    \end{subfigure}
    \qquad
    \begin{subfigure}[b]{.95\textwidth}
        \includegraphics[width=\linewidth]{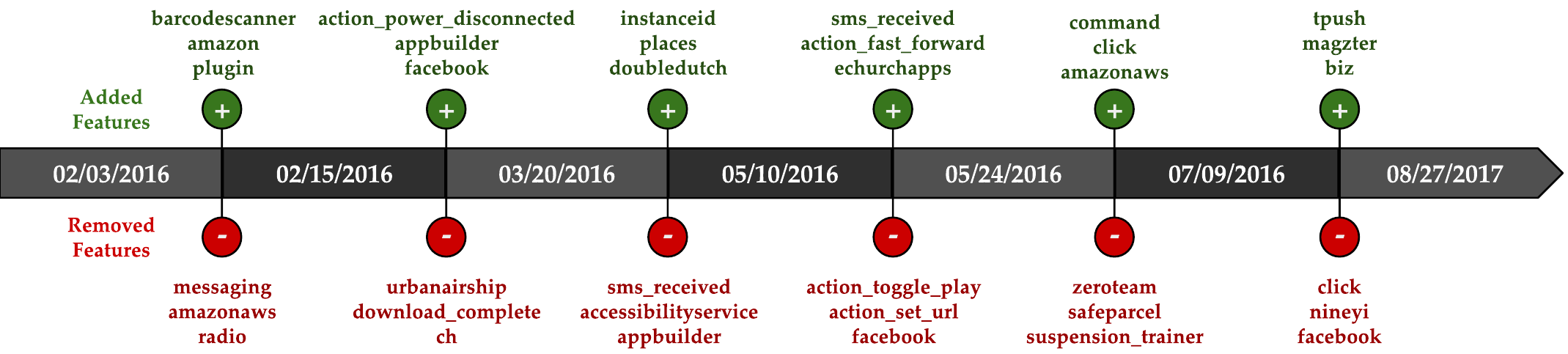}
        \caption{Vocabulary change timeline for manifest and resources
        using AndroZoo dataset.}
        \label{fig:androzoo_vocabulary_change}
    \end{subfigure}
    
    \caption{\textbf{Vocabulary changes for both datasets.} Many significant features are removed and added as time goes by.}
    \label{fig:vocabulary_change}
\end{figure*}

\noindent \textit{DREBIN Dataset Evolution.}
Figure~\ref{fig:drebin_vocabulary_change}
shows API calls' vocabulary change for the first six
detected drift points 
that actually presented 
changes (green and red words mean
introduced and removed from the vocabulary, respectively).
We identified periodic trends and system evolution as the two main reasons for the constant change
of malware samples.
The first occurs because malware creators
compromise systems according to the available infection vectors (e.g., periodic events exploitable via social engineering). Also, attackers usually shift their operations to distinct
targets when their infection strategies become so popular that AVs detect them. Therefore, some
features periodically enter and leave the vocabulary
(e.g., \texttt{setcontentview}). 
The second occurs due to changes in the Android platform, causing reactions from
 malware creators to handle the new scenario, either by supporting
newly introduced APIs 
(e.g., \texttt{getauthtoken}~\cite{authtoken}), 
or by unsupporting deprecated and modified
APIs (e.g., deleted \texttt{keyguard}~\cite{keyguard}
or the deprecated \texttt{fbreader}~\cite{fbr} intent).
Handling platform evolution is required to 
keep malware samples working on newer systems, whereas ensuring maliciousness. In this sense,
the removal of a feature like \texttt{DELETE\_PACKAGE} 
permission from the vocabulary can be explained by
the fact that Android changed permission's transparency 
behavior to an explicit confirmation request~\cite{deleteperm}, 
which makes malware abusing those features less effective.
The most noticeable case of malware evolution involves
SMS sending, a feature known to be abused by
plenty of samples~\cite{Sarma:2012:APP:2295136.2295141,Hamandi:2012:ASB:2386995.2387016,Luo:2013:RDP:2484417.2484422}.
We identified that APIs were periodically 
added and removed from the vocabulary (e.g., \texttt{sendmultiparttextmessage}, part of
the Android SMS subsystem~\cite{sms1}, had its use restricted by Android over time, until being finally blocked~\cite{smsblock}). This will certainly cause modifications in newer
malware and probably incur in classifiers'
drifting once again. Therefore, considering
concept drift is essential to define any realistic
threat model and malware detector. 
%
%

\noindent \textit{AndroZoo Dataset Evolution.}
We conducted the same experiments performed on the Drebin dataset on the
AndroZoo dataset to evaluate dataset influence on concept drift occurrence, as shown in Figure~\ref{fig:androzoo_vocabulary_change}.
We highlight that developing a malware classifier for the AndroZoo dataset
is a distinct challenge than developing a malware classifier for the Drebin
dataset because the AndroZoo dataset presents more malware families (1.166
distinct families according to Euphony~\cite{7962391} labels) than the Drebin
dataset (178 distinct families according to Euphony labels). The difference in
the dataset complexity is reflected in the number of drift points
identified for each experiment. Whereas we identified 18 drift points for
the Drebin dataset, the AndroZoo dataset presented 50 drift points. 
Despite the difference in the number of drifting points, the reasons behind such
drifts have been revealed very similar when we delve into vocabulary change
details. When we look to the \texttt{manifest\_action} vocabulary, we
notice that the \texttt{sms\_received} feature leaves and returns to the
vocabulary many times. This behavior reflects the arms race between Google
and attackers for SMS sending permission~\cite{smsblock}. This same occurrence has been
observed in the DREBIN dataset. Similarly, when we look to the
\texttt{manifest\_category} vocabulary, we notice that the
\texttt{facebook} feature also leaves and returns to the vocabulary many
times. This happens because the attackers often rely on Facebook name for
distributing ``Trojanized'' applications that resemble that original
Facebook app. Although Google often removes these apps from the official
store, attackers periodically come up with new ways of bypassing AppStore's
verification routines, thus resulting in the presented infection waves
regarding the Facebook name. Finally, this same behavior is observed in the
\texttt{resource\_entry} vocabulary regarding the \texttt{amazonaws}
feature. Attackers often store their malicious payloads on popular cloud
servers to avoid detection~\cite{10.1007/978-3-642-37300-8_3}. Even though Amazon sinkhole malicious domains
when detected, attackers periodically find new ways to survive security
scans, thus also resulting in the observed vocabulary waves.

The results of our experiments show that despite presenting distinct
complexities, both datasets have undergone 
drift occurrences.
This shows that concept drift is not a dataset-exclusive issue, but a
characteristic inherently associated with the malware classification
problem. Therefore, overall malware classification research work (and
primarily the ones leveraging these two popular datasets) should consider
concept drift occurrence in their evaluations to gather more accurate
samples information.

\section{Discussion}
\label{discussion}

We here discuss our findings and their implications.

\noindent \textbf{What the metrics say.}
While the accuracy
of distinct classifiers is very
similar in 
some experiments, we highlight that
malware detectors 
must be evaluated
using the correct metrics, as 
any binary classification. 
Therefore, we 
rely on f1score, recall,
and precision
to gain further knowledge on malware 
detection. AV companies should adopt the same reasoning when evaluating their classifiers. 
Experiments~\ref{exp2}
and~\ref{exp3} indicate Android malware evolution, which imposes
difficulties for established classifiers and corroborates the 
fact that samples 
changed. 
However, goodware 
generally kept the same concept
over time, suggesting that creating benign profiles may be better than modeling everchanging malicious activities.

\noindent \textbf{Feature drift, concept drift and evolution.}
Using a fixed ML model is not enough
to predict future threats. In 
experiment~\ref{exp2}, we notice a
significant drop in f1score and recall 
(a metric that indicates the ability 
to correctly classify malware). 
Hence, a fixed ML model is prone to
misclassify novel threats.
Besides, continuous learning helps
increasing detection but is still 
subject to drift, as shown by AndroZoo 
in experiment~\ref{exp3}, when recall remains below 10\% for
a long period.
If we compare experiments~\ref{exp3} and~\ref{exp4} when classifying DREBIN, 
it is possible to observe that the Incremental Windowed Classifier update 
still outperforms DDM with the update when \review{using Adaptive Random Forest}, 
but is not better than DDM with retraining 
(except in precision). However, when 
comparing it with EDDM and ADWIN, we notice that both are significantly better, and it is still highly prone to drift (Figure~\ref{fig:exp3}).
In response to these results, concept drift 
detectors help to improve classification performance.
When comparing DDM with retraining and EDDM
and ADWIN (with both update and retrain) to 
IWC (Table~\ref{table:results}), we can
see that the former outperformed the latter,
advocating the need of using 
drift detectors. \review{This is evident when using SGD, 
once IWC was outperformced by all other methods, and}
is even more evident when classifying AndroZoo, 
given that
IWC lost the ability to detect new malware as time goes by. In this case,
ADWIN with retraining was able to outperform all the other methods again, even a closely related work (DroidEvolver~\cite{Xu2019}), giving us insight that this is a valid method to be used in practice. 
In addition, even more important than just using a drift detector,
reconsidering the classifier's 
feature set
is required to
completely overcome concept drift.
According to experiment~\ref{exp4}, we can infer 
that retraining the classifier and
feature extraction models every time 
a drift occurs (using the data stream 
pipeline we proposed) is better than just 
updating only the classifier. This implies 
that not only the representation of malware
becomes obsolete as time goes by, but also 
the vocabulary used to build them. It means
that 
every textual attribute
used by applications can change, i.e., new API
calls, permissions, URLs, 
for example,
emerge, requiring a vocabulary update, i.e.,
%
it indicates that discovering new specific 
features might help in increasing 
detection rates. Thus, we conclude that malware detection
is not just a concept drift problem, but also in essence a
feature drift detection problem~\cite{BARDDAL2017278}.

\noindent \textbf{Our solution in practice.}
To implement our solution in practice, we need to model
the installation, update, and removal of applications as a stream.
This can be done by considering the initial set of applications
installed in a stock phone as ground truth and thus subsequent
deployment of applications as the stream. To reduce the
delay between the identification of a drift point and the 
classifier update, ideally, multiple classifiers (the current
one and the candidates to replace it) should be running in
parallel. However, this parallel execution is too much
expensive to be performed on endpoints. Therefore, we consider
that the best usage scenario for our approach is its deployment
on the App's Stores distributing the applications. Therefore,
each time a new application is submitted to an App Store, it is
verified according to our cycle. To cover threats distributed
by alternative markets, this concept can be extended to any
centralized processing entity. For instance, an AV can
upload the suspicious application to a cloud server and
perform its checks according to our cycle. To speed up
the upload process, the AV can make the current feature
extractor available to the endpoint so as the endpoint does
not need to upload the entire application but only its feature.
In this scenario, the AV update is not composed of new signatures,
but of new feature extractors.

\noindent \textbf{Drift and evolution are common problems in malware detection.}
Despite 
being more complex,
(more apps and 
malware families) 
all 
characteristics present in DREBIN 
are drastically shown in AndroZoo, 
evidencing that feature drift, concept drift, and evolution are
present in malware detection problems in practice
and not only in a single dataset.
This suggests that AV companies should keep
enhancing their models and knowledge database
constantly. 

\noindent \textbf{Limitations and future work.}
One of the limitations of our 
approach is that it relies on ground truth labels that may
be available with a certain delay, given that known goodware
and malware are needed in order to train and update the 
classifier (as any AV). Thus, future researches to reduce 
these delays are an important step to improve any ML
solution that does not rely on their own labels, such as 
DroidEvolver~\cite{Xu2019} (outperformed by our 
approach).

\reviewtwo{It is important to note that our work considers only the attributes vectors provided by DREBIN~\cite{arp2014} and AndroZoo~\cite{androzoo} datasets' authors. Due to this fact, we were unable to consider other types of attributes, such as API semantics~\cite{10.1145/3372297.3417291} or behavioral profiling (dynamic analysis)~\cite{10.1145/3371924}. Therefore, we cannot directly compare our work with theirs, due to different threat models. To compare different attributes using our data stream pipeline under the same circumstances, it is necessary to download all the APKs from both datasets. This is left as future work.}




Finally, we make our data stream 
learning pipeline available as an extension of 
\texttt{scikit-multiflow}~\cite{skmultiflow}, aiming at encouraging other researchers to contribute with new 
strategies that address feature and concept drift. 

\section{Conclusion}
\label{conclusion}

In this article, we evaluated the impact of concept drift on malware classifiers for 
Android malware samples to understand how fast 
classifiers expire. We analyzed $\approx$415K sample Android apps from two datasets (DREBIN and AndroZoo) collected over nine years (2009-2018) using two 
representations (Word2Vec and TF-IDF), \review{two classifiers (Adaptive Random Forest and Stochastic Gradient Descent classifier) and four 
drift detectors (DDM, EDDM, ADWIN, and KSWIN)}. Our results show that resetting the classifier only after changes are detected is better than periodically resetting it based on a fixed window length. We also point the need to update the feature extractor (besides the classifier) to achieve increased detection rates, due to new features that may appear over time. This strategy was the best in all scenarios presented, even using a 
complex dataset, such as AndroZoo. The implementation is available as an extension to \texttt{scikit-multiflow}~\cite{skmultiflow}\footnote{\url{https://github.com/fabriciojoc/scikit-multiflow}}. Our results highlight the need for developing new strategies to update the classifiers inherent to AVs.


\bibliography{refs}

\end{document}